\documentclass[journal=jacsat,manuscript=article,layout=twocolumn]{achemso}
\setkeys{acs}{articletitle = true}

\usepackage[version=3]{mhchem} 
\usepackage{amssymb,color,amsmath,csvsimple,tabularx}
\usepackage[utf8]{inputenc}
\usepackage{subfiles}
\usepackage{numprint}
\usepackage{makecell}
\usepackage{adjustbox}

\newcommand{\pipeline}{AMPL }

\author{Amanda J. Minnich}
\email{minnich2@llnl.gov}
\author{Kevin McLoughlin}
\affiliation{Lawrence Livermore National Laboratory}
\author{Margaret Tse}
\author{Jason Deng}
\author{Andrew Weber}
\author{Neha Murad}
\affiliation{GlaxoSmithKline}
\author{Benjamin D. Madej}
\affiliation{Frederick National Laboratory}
\author{Bharath Ramsundar}
\affiliation{Computable}
\author{Tom Rush}
\author{Stacie Calad-Thomson}
\affiliation{GlaxoSmithKline}
\author{Jim Brase}
\author{Jonathan E. Allen}
\affiliation{Lawrence Livermore National Laboratory}

\title{AMPL: A Data-Driven Modeling Pipeline for Drug Discovery}

\abbreviations{IR,NMR,UV}

\begin{document}
\maketitle
\begin{abstract}
    
One of the key requirements for incorporating machine learning into the drug discovery process is complete reproducibility and traceability of the model building and evaluation process. With this in mind, we have developed an end-to-end modular and extensible software pipeline for building and sharing machine learning models that predict key pharma-relevant parameters. The ATOM Modeling PipeLine, or AMPL, extends the functionality of the open source library DeepChem and supports an array of machine learning and molecular featurization tools. We have benchmarked AMPL on a large collection of pharmaceutical datasets covering a wide range of parameters.  Our key findings include:
\begin{itemize}
    \item Physicochemical descriptors and deep learning-based graph representations significantly outperform traditional fingerprints in the characterization of molecular features. 
    \item Dataset size is directly correlated to prediction performance, and that single-task deep learning models only outperform shallow learners if there is sufficient data. Likewise, dataset size has a direct impact on model predictivity, independent of comprehensive hyperparameter model tuning. Our findings point to the need for public dataset integration or multi-task/transfer learning approaches. 
    \item Uncertainty quantification (UQ) analysis may help identify model error; however, efficacy of UQ to filter predictions varies considerably between datasets and featurization/model types. 
    
    \end{itemize}
AMPL is open source and available for download at \url{http://github.com/ATOMconsortium/AMPL}.

\end{abstract} \hspace{10pt}

\section{Introduction}

Discovery of new compounds to treat human disease is a multifaceted process involving the selection of chemicals with favorable pharmacological properties: a high potency to the desired target, elimination or minimization of safety liabilities, and a favorable pharmacokinetic (PK) profile. To address this challenge, the drug discoverer has a wealth of choices, with total ``drug-like” chemical matter estimated between $10^{22}$-$10^{60}$ unique molecules. However, evaluating the desirability of these molecules with respect to potency, pharmacokinetics, and safety liabilities is a time-consuming and expensive process. Many of these molecules require \textit{de novo} synthesis, which is a rate-limiting step. Furthermore, evaluation of pharmacological properties both \textit{in vitro} and especially \textit{in vivo} is prohibitively expensive given the universe of possible choices. 

To aid in this design challenge, the field of computer-aided drug design has evolved to rapidly predict the properties of pharmacological matter \textit{in silico}, allowing for rational selection of a feasible set compounds for synthesis and evaluation. These techniques generally fall into two categories: (1) structure-based drug design, which relies on knowledge of the target structure (i.e. docking, molecular dynamics, free energy perturbation) and (2) ligand-based drug design, which uses known properties of molecules to develop models of quantitative structure-activity relationships (QSAR).

Ligand-based drug design generally relies on machine learning-based techniques to identify the link between structure and the property of interest. Recently, a proliferation of advanced machine learning techniques have shown great promise in increasing the predictability of QSAR models. A deep learning model first won the Merck Kaggle multi-activity challenge in 2014\cite{dahl_multi-task_2014}, and since then these models have continued to show increased predictive accuracy over QSAR models based on classical machine learning techniques in many studies\cite{gilmer_neural_2017}. A recent example of success with deep learning is the paper by Feinberg et al. that compared the PotentialNet deep learning method with existing shallow learners on a wide array of pharmaceutically-relevant datasets \cite{feinberg_step_2019}.  These results showed dramatic improvements for deep learning based on temporal splits using data collected from a pharmaceutical company. Another evaluation showed that a directed message-passing neural network model can provide robust performance over a range of experimental datasets characterizing molecular properties\cite{yang_analyzing_2019}. The authors provide an open-source deep learning software to go with this paper that has been tested on a wide range of parameters. However, this software does not include any type of modular pipeline that would allow for the incorporation of different models and chemical representations. Overall there has been a lack of publicly available suites of software tools that support a transparent and reproducible generation of a diverse array of deep and classical machine learning models, especially ones that can scale to model the large set of pharmaceutically-relevant parameters. A major advance towards this goal was made with the introduction of DeepChem \cite{deepchem}, which supports the building of a variety of machine learning models for small molecule property prediction. DeepChem contains a variety of very helpful modules and tools, but has limitations in its ability to robustly train models from a wide selection of hyperparameters, and published performance evaluation is limited to a small number of public datasets with less diverse pharmaceutical relevance \cite{wu_moleculenet:_2018}. 

In this paper we introduce a new small molecule property prediction pipeline, AMPL.  This software was developed through the Accelerating Therapeutics Opportunities in Medicine (ATOM) Consortium as the ATOM Modeling PipeLine. The key contributions of this work are to automate deep learning training, particularly in hyperparameter search; to enable extensive performance benchmarking; and to apply \pipeline to a large collection of pharmaceutically-relevant property-prediction datasets. Most notably, \pipeline is available as open source to benefit the drug discovery community.

The closest existing pipeline tools are BIOVIA Pipeline Pilot \cite{pipeline_pilot} and KNIME \cite{knime}. Pipeline Pilot is a license-based graphical tool for machine learning pipelining. It has capabilities for data cleaning, splitting, training, and model deployment, but are all mainly GUI-based, limiting the customizability of the software. Furthermore, it is only available for a licensing fee, so it does not target the open source community. In terms of free and open source software suites for data analytics, the main alternative is KNIME. This software provides an environment for creating general data flows to process data, use predictive models, and analyze complex datasets. An ecosystem of open source and commercial KNIME node extensions has developed which enable workflows for library analyses, virtual screening, model fitting and prediction. In contrast, \pipeline is tightly focused on integrating modern machine learning methods with best practices for chemical activity and property prediction. Important issues with machine learning models, such as dataset characterization, model validation, and uncertainty quantification are addressed by \pipeline in automated and reproducible ways. The code suite also provides high performance computing modules for model fitting, hyperparameter optimization, and predictions. AMPL currently targets job submission-based clusters to scale training runs; however, AMPL could be adapted to operate on other scalable platforms such as Spark in the future. Furthermore, \pipeline  is implemented as a modular and reusable Python library to allow for easy integration with other data science software platforms.

An extensive set of experiments were conducted with AMPL, and key observations include: \begin{itemize} 
    \item Physicochemical descriptors and deep learning-based graph representations are significantly better than traditional fingerprints to characterize molecular features
    \item Dataset size is directly correlated to performance of prediction: single-task deep learming models only outperform shallow learners if there is enough data. Likewise, data set size has a direct impact of model predictivity independently of comprehensive hyperparameter model tuning. Our findings point to the need for public dataset integration or multi- task/ transfer learning approaches.
    \item DeepChem uncertainty quantification (UQ) analysis may help identify model error; however,  efficacy of UQ to filter predictions varies considerably between datasets and model types.
\end{itemize}

The aim of this paper is to present the rigorous and transparent open source software pipeline \pipeline to build global and local `baseline’ models for a wide array of molecular properties that are needed for \textit{in silico} drug discovery. This new software will support reproducible training and testing protocols that enable the broader modeling community to evaluate and improve modeling approaches over time.

\section{Methods}

\begin{figure*}
    \centering
    \includegraphics[width=.9\textwidth]{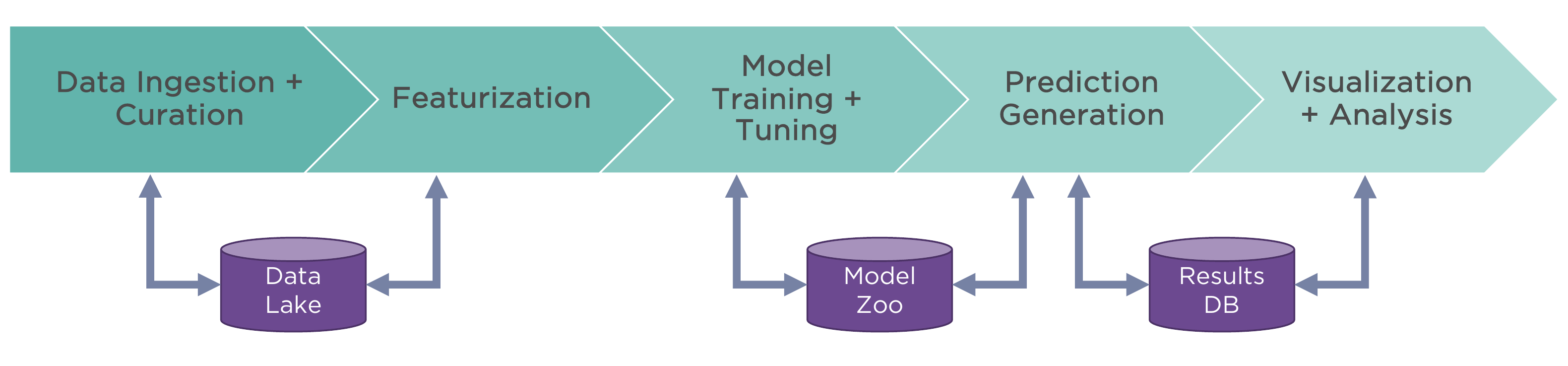}
    \caption{Overview of \pipeline}
    \label{fig:pipeline}
\end{figure*}

Figure \ref{fig:pipeline} shows the overall architecture of AMPL. This end-to-end pipeline supports all functions needed to generate, evaluate, and save machine learning models: data ingestion and curation, featurization of chemical structures into feature vectors, training and tuning of models, storage of serialized models and metadata, and visualization and analysis of results. It also contains modules for parallelized hyperparameter search on high-performance computing (HPC) clusters.

\subsection{Data curation}
 \pipeline includes several modules to curate data into machine learning-ready datasets. Functions are provided to represent small molecules with canonicalized SMILES strings using RDKit \cite{landrum_fingerprints_nodate} and the MolVS package \cite{matt_swain_2017_260237}, by default stripping salts and preserving isomeric forms. Data curation procedures are provided with \pipeline as Jupyter notebooks \cite{noauthor_jupyterlab_nodate}, which can be used as examples for curating new datasets.  Procedures allow for averaging response values for compounds with replicate measurements and filtering compounds with high variability in their measured response values. \pipeline also provides functions to assess the structural diversity of the dataset, using either Tanimoto distances between fingerprints, or Euclidean distances between descriptor feature vectors. 
 
Data ingestion and curation-related parameters include:
 \begin{itemize}
     \item Unique human readable name for training file
     \item Data privilege access group
     \item Parameter for overriding the output files/dataset object names
     \item ID for the metadata + dataset
     \item Boolean flag for using an input file from the file system
     \item Name of column containing compound IDs 
      \item Name of column containing SMILES strings
     \item List of prediction task names 
     \item Number of classes for classification
     \item User specified list of names of each class
     \item Boolean switch for using transformation on regression output. Default is True
      \item Response column normalization type 
     \item Minimum number of dataset compounds considered adequate for model training. A warning message will be issued if the dataset size is less than this. 
 \end{itemize}
\subsection{Featurization}

\pipeline provides an extensible featurization module which can generate a variety of molecular feature types, given SMILES strings as input. They include:

\begin{itemize}
\item Extended connectivity fingerprints (ECFP) with arbitrary radius and bit vector length \cite{rogers_extended-connectivity_2010}
\item Graph convolution latent vectors, as implemented in DeepChem \cite{duvenaud_convolutional_2015}
\item Chemical descriptors generated by the Mordred open source package \cite{moriwaki_mordred:_2018}
\item Descriptors generated by the commercial software package Molecular Operating Environment (MOE) \cite{noauthor_chemical_nodate}
\item User-defined custom feature classes
\end{itemize}

Because some types of features are expensive to compute, \pipeline supports two kinds of interaction with external featurizers: a dynamic mode in which features are computed on-the-fly and a persistent mode whereby features are read from precomputed tables and matched by compound ID or SMILES string. In the persistent mode, when SMILES strings are available as inputs, the featurization module matches them against the precomputed features where possible, and computes features dynamically for the remainder. Because precomputed feature tables may span hundreds or thousands of feature columns for millions of compounds, the module uses the feather format \cite{noauthor_feather_nodate} to speed up access.

Featurized datasets for feature types that support persistent mode (currently, all except ECFP fingerprints and graph convolution format) are saved in the filesystem or remote datastore, so that multiple models can be trained on the same dataset. This also facilitates reproducibility of model results.

Chemical descriptor sets such as those generated by MOE often contain descriptors that are exact duplicates or simple functions of other descriptors. In addition, large blocks of descriptors may be strongly correlated with one another, often because they scale with the size of the molecule. The featurization module deals with this redundancy by providing an option to remove duplicate descriptors and to scale a subset of descriptors by the number of atoms in the molecule (while preserving the atom count as a distinct feature). Factoring out the size dependency often leads to better predictivity of models.

The featurization module can be easily extended to handle descriptors generated by other software packages, latent vectors generated by autoencoders, and other types of chemical fingerprints. In most cases, this can be accomplished by writing a small function to invoke the external feature generation software, and by adding an entry to a table of descriptor types, listing the generated feature columns to be used. In more complicated cases, one may need to write a custom subclass of one of the base featurization classes. 

Featurization-relevant input parameters include:

\begin{itemize}
    \item Type of molecule featurizer
    \item Feature matrix normalization type   
    \item Boolean flag for loading in previously featurized data files
    \item Type of transformation for the features
    \item Radius used for ECFP generation
    \item Size of ECFP bit vectors
    \item Type of autoencoder, e.g. molvae, jt
    \item Trained model HDF5 file path, only needed for MolVAE featurizer
    \item Type of descriptors, e.g. MOE, Mordred 
    \item Max number of CPUs to use for Mordred descriptor computations. None means use all available
    \item Base of key for descriptor table file
\end{itemize}

\subsection{Dataset partitioning}
\pipeline supports several options for partitioning datasets for model training and evaluation,  By default, datasets are split into 3 parts: a training set, a validation set (for parameter selection), and a holdout test set (for evaluation). Alternatively, \pipeline offers a k-fold cross-validation option, to assess the performance impact of sampling from the modeled dataset. Under k-fold cross-validation, the holdout test set is selected first, and the remainder is divided into k-fold sets for training and validation.

\pipeline offers a number of dataset splitting algorithms, which offer different approaches to the problem of building models that generalize from training data to novel chemical space. It supports several of the methods included in DeepChem, including random splits, Butina clustering, Bemis-Murcko scaffold splitting, and a simple algorithm based on fingerprint dissimilarity \cite{wu_moleculenet:_2018}. In addition, we implemented temporal splitting and a modified version of the asymmetric validation embedding (AVE) debiasing algorithm \cite{wallach_most_2018}. We compared random splitting with Bemis-Murcko scaffold splitting for our benchmarking experiments.

Input parameters related to data splitting include:
\begin{itemize}
    \item Type of splitter to use: index, random, scaffold, Butina, ave\_min, temporal, fingerprint, or stratified
    \item Boolean flag for loading in previously-split train, validation, and test CSV files
    \item UUID for CSV file containing train, validation, and test split information
    \item Choice of splitting type between k-fold cross validation and a normal train/valid/test split
    \item Number of k-folds to use in k-fold cross validation
    \item Type of splitter to use for train/validation split if temporal split used for test set (random, scaffold, or ave\_min)
    \item Cutoff Tanimoto similarity for clustering in Butina splitter
    \item Cutoff date for test set compounds in temporal splitter
    \item Column in dataset containing dates for temporal splitter
    \item Fraction of data to put in validation set for train/valid/test split strategy 
    \item Fraction of data to put in held-out test set for train/valid/test split strategy 
\end{itemize}                                         
\subsection{Model training and tuning}
\pipeline includes a train/tune/predict framework to create high-quality models. This framework supports a variety of model types from two main libraries: scikit-learn \cite{sklearn} and DeepChem \cite{deepchem}.  Currently, specific input parameters are supported for:
\begin{itemize}
    \item Random forest models from scikit-learn
    \item XGBoost models \cite{xgboost}
    \item Fully connected neural network models
    \item Graph convolution neural network models \cite{graphconv}
\end{itemize} As with the featurization module, \pipeline supports integration of custom model sub-classes. Parameters for additional models can be easily added to the parameter parser module. 

Model-relevant input parameters include:
\begin{itemize}
 \item Type of model to fit (neural network, random forest, or xgboost)
 \item Prediction type (regression or classification)
 \item Singletask or multitask model 
 \item Number of decision trees in the forest for random forest models  
 \item  Max number of features to split random forest nodes
 \item Number of estimators to use in random forest models
 \item Batch size for neural network model
 \item Optimizer type for neural network model
 \item Optimizer specific for graph convolutional models, defaults to ``adam"
 \item Model batch size for neural network model
 \item List of hidden layer sizes for neural network model 
 \item List of dropout rates per layer neural network model
 \item List of standard deviations per layer for initializing weights for neural network model  
 \item The type of penalty to use for weight decay, either ``l1" or ``l2"
 \item The magnitude of the weight decay penalty to use
 \item List of initial bias parameters per layer for neural network model 
 \item Learning rate for dense neural network models 
 \item Epoch for evaluating baseline neural network model performance, if desired
 \item Maximum number of training epochs for neural network model             
 \item Type of score function used to choose best epoch and/or hyperparameters
 \item Boolean flag for computing uncertainty estimates for regression model predictions
\end{itemize} 
\subsubsection{Epoch selection for neural network models}
Early stopping is an essential strategy to avoid overfitting of neural networks, thus the number of training epochs is one of the key hyperparameters that must be optimized. To implement early stopping, \pipeline trains neural network models for a user-specified maximum number of epochs, evaluating the model on the validation set after each epoch, and identifies the epoch at which a specified performance metric is maximized. By default this metric is the coefficient of determination $R^2$ for regression models, and the area under the receiver operating characteristic curve (ROC AUC) for classification models.
\subsubsection{Model persistence}
Serialized models are saved after training and prediction generation are complete, along with detailed metadata to describe the model. This supports traceability and reproducibility, as well as model sharing. \pipeline supports saving models and results either using the file system or optionally through a collection of database services. The metadata can be stored in a mongoDB database \cite{mongo} or as JSON files. \pipeline has functions for saving models and loading pre-built models for prediction generation. 

\subsection{Model performance metrics}
\pipeline calculates a variety of performance metrics for predictions on the training, validation and test sets. Metrics may be saved in a mongoDB database or in JSON files. 
For regression models, we calculate:
\begin{itemize}
    \item Coefficient of determination ($R^2$). This is calculated using sklearn's metrics function. Note that this score can be negative if the model is arbitrarily worse than random.
    \begin{equation}
        R^2(y, \hat{y}) = 1 -\frac{\sum_{i=1}^{n} (y_i - \hat{y_i})^2}{\sum_{i=1}^{n} (y_i - \bar{y})^2}
    \end{equation}
    \item Mean Absolute Error (MAE).  An advantage of MAE is that it has a clear interpretation, the average absolute difference between the measured value $y_i$ and predicted value $\hat{y_i}$. This works well for cellular activity assay datasets, which use log normalized dose concentration value with similar concentration ranges across different assays. PK parameters are measured on different scales for some assays, which prevents comparison between assays with this metric.
    \begin{equation}
        \mathrm{MAE} = \frac{\sum_{i=1}^{n} |y_i - \hat{y_i}|}{n}
    \end{equation}
    \item Mean Square Error (MSE). This is a risk metric corresponding to the expected value of the squared error (or loss). 
    \begin{equation}
        \mathrm{MSE}(y, \hat{y}) = \frac{1}{n} \sum_{i=0}^{n - 1} (y_i - \hat{y_i})^2
    \end{equation}
\end{itemize}

For classification models, we calculate:
\begin{itemize}
    \item Area Under the Receiver Operating Characteristics Curve (ROC AUC). The ROC curve plots the True Positive Rate versus the False Positive Rate as a binary classifier's discrimination threshold is varied. The ROC AUC score is calculated by finding the area under the ROC Curve. This value can range from 0 -- 1, where 1 is the best score.
    \item Precision (Positive Predictive Value)
    \begin{equation}
        \mathrm{Precision} = \frac{TP}{TP+FP}
    \end{equation}
    where TP = number of true positives and FP = number of false positives
    \item Recall (True positive rate/ sensitivity)
    \begin{equation}
        \mathrm{Recall} = \frac{TP}{TP+TN}
    \end{equation}
    where TP = number of true positives and TN = number of true negatives
    \item Area under the precision-recall curve (PRC-AUC). The precision-recall curve plots precision versus recall as a binary classifier's discrimination threshold is varied. It is a good measure of success of prediction when classes are very imbalanced. High scores show that the classifier is returning accurate results (high precision), as well as returning a majority of all positive results (high recall).
    \item Negative Predictive Value (NPV)
    \begin{equation}
        \mathrm{NPV} = \frac{TN}{TN+FN}
    \end{equation}
    where TN = number of true negatives and FN = number of false negatives
    \item Cross entropy (log loss)
    \begin{equation}
        \mathrm{Cross\ entropy} = -\sum_{c=1}^{M}{y_{o,c}log(p_{o,c})}
    \end{equation}
    \item Accuracy
    \begin{equation}
        \mathrm{Accuracy} = \frac{TP+TN}{TP+TN+FP+FN}
    \end{equation}
    where terms are defined as above.
\end{itemize}

\subsection{Uncertainty quantification}
Uncertainty quantification (UQ) attempts to measure confidence in a model's prediction accuracy by characterizing variance in model predictions. Some common objectives for UQ are to use it to guide active learning or to weight model ensembles. \pipeline generates UQ values for both random forest and neural network models.
\subsubsection{Uncertainty quantification for random forest}
Generating a value quantifying uncertainty is straightforward for random forest and is taken to be the standard deviation of predictions from individual trees. This quantifies how variable these predictions are, and thus how uncertain the model is in its prediction for a given sample.
\subsubsection{Uncertainty quantification for neural networks}
Our neural network-based UQ uses the Kendall and Gal method\cite{kendall_what_2017} as implemented in DeepChem. This method combines aleatoric and epistemic uncertainty values. 

Aleatoric uncertainty cannot be reduced by adding more data but can be estimated. It is estimated by modifying the loss function of the model to predict both the response variable and the error of the model. 

Epistemic uncertainty arises because of limited data. It represents the uncertainty of the model. Normally this is calculated in a bootstrapped manner, as in the case of a random forest. However, since training neural networks is expensive, an alternate approach is to train one network to generate a set of predictions by applying a set of dropout masks during prediction. Prediction variability is then quantified to assess epistemic uncertainty. 

\subsection{Visualization and analysis}
Plots generated by AMPL's visualization and analysis module are shown in the Results section. Additional options include plots of predicted vs. actual values, learning curves, ROC curves, precision vs. recall curves, and 2-D projections of feature vectors using UMAP \cite{mcinnes_umap_2018}. The module also includes functions for characterizing and visualizing chemical diversity. Chemical diversity analysis is crucial for analyzing domain of applicability, bias in dataset splitting, and novelty of \textit{de novo} compounds. This module supports a wide range of input feature types, distance metrics, and clustering methods.

\subsection{Hyperparameter optimization}
A module is available to support distributed hyperparameter search for HPC clusters. This module currently supports linear grid, logistic grid, and random hyperparameter searches, as well as iteration over user-specified values. To run the hyperparameter search, the user specifies the desired range of configurations in a JSON file. The user can either specify a single dataset file or a CSV file with a list of datasets. The script generates all valid combinations of the specified hyperparameters, accounting for model and featurization type, and submits jobs for each combination to the HPC job scheduler. The module includes an option to generate a pre-featurized and pre-split dataset before launching the model training runs, so that all runs operate on the same dataset split. The user can specify a list of layer sizes and dropouts to combine, along with the maximum final layer size and a list of the numbers of possible layers for a given model, and the module combines these different options based on the input constraints to generate a variety of model architectures. The search module can check the model tracker database to avoid retraining models that are already available. It also provides users the option to exclude hyperparameter combinations that lead to overparameterized models, by checking the number of weight and bias parameters for a proposed neural network architecture against the size of the training dataset. Finally, the search module throttles job submissions to prevent the user from monopolizing the HPC cluster. 

Input parameters for hyperparameter search include:
\begin{itemize}
     \item Boolean flag indicating whether we are running the hyperparameter search script
     \item UUID of hyperparam search run model was generated in
     \item Comma-separated list of number of layers for permutation of NN layers
      \item Comma-separated list of dropout rates for permutation of neural network layers
     \item The maximum number of nodes in the last layer
     \item Comma-separated list of number of nodes per layer for permutation of neural network layers
     \item Maximum number of jobs to be in the queue at one time for an HPC cluster
     \item Scaling factor for constraining network size based on number of parameters in the network
     \item Boolean flag directing whether to check model tracker to see if a model with that particular param combination has already been built
     \item Path where pipeline file you want to run hyperparam search from is located 
     \item Type of hyperparameter search to do. Options are grid, random, geometric, and user\_specified 
     \item CSV file containing list of datasets of interest
\end{itemize}

\subsection{Running \pipeline}
There are three ways to run AMPL:
\begin{itemize}
    \item Using a config file: Create a JSON file with desired model parameters and run full pipeline via command line
    \item Using command line arguments: Specify model parameters via standard command line arguments
    \item Interactively in a Jupyter notebook using an argparse.Namespace object or a dictionary
\end{itemize}

\section{Results}\label{sec:results}

Benchmark experiments were run to evaluate and validate components of the pipeline.

\subsection{Data}

Experimental datasets were made available by ATOM Consortium member GlaxoSmithKline from a variety of bioactivity and pharmacokinetics experiments. Selected datasets were used for training and evaluating models. 
These datasets are summarized in Table
\ref{tab:PharmSum}.

 Pharmacokinetic (PK) and safety datasets were curated separately, as they contain different types of experimental data and thus require different processing.
 The raw datasets were cleaned to remove rows with outlying, missing, and duplicate measurements, and processed to yield machine learning datasets with a single aggregate value per unique compound.  These procedures informed the design of curation functions included in the pipeline. Curation of the PK datasets required the conversion of values to standard units, the removal of compounds with stability or recovery issues, and the exclusion of data that was generated using significantly different assay protocols.  Subsequently, replicate experimental measurements were identified by matching duplicate canonical SMILES strings and averaged to produce a single value per compound.
 
 For the safety datasets, censored measurements were an additional concern. Since bioactivity assays are typically performed over a limited range of compound concentrations, IC50 or EC50 values may be reported as being above or below a maximum or minimum concentration, so that the measurements are censored.  When all measurements for a compound are censored in the same direction, the user is given the option to either exclude the compound from the dataset, or include it with a relational operator indicating the direction together with the censoring threshold. In the case where some replicate measurements are censored and some are not, \pipeline computes a maximum likelihood estimate for the mean activity, assuming a Gaussian distribution of measurements around the true mean. The distribution of response values is reported along with the mean and standard deviation. 

\begin{table*}[ht]
\caption{Pharmacokinetics datasets used to benchmark \pipeline}
\resizebox{\textwidth}{!}{%
\begin{tabular}{|l|l|l|l|l|l|l|l|}
\hline
\textbf{Dataset} & \textbf{Units} & \textbf{Species} & \textbf{Dataset Size} & \textbf{Minimum} & \textbf{Maximum} & \textbf{Mean} & \textbf{Median} \\ \hline
Blood to Plasma Ratio &  & Human & 101 & 0.47 & 10.5 & 0.85 & 0.77 \\ \hline
Blood to Plasma Ratio &  & Dog & 71 & 0.37 & 6.85 & 0.85 & 0.88 \\ \hline
Plasma Protein Binding HSA & fraction unbound & Human & 123734 & 0.0001 & 1 & 0.05 & 0.044 \\ \hline
Plasma Protein Binding HSA & fraction unbound & Rat & 2086 & 0.0001 & 1 & 0.036 & 0.033 \\ \hline
Plasma Clearance (\textit{In Vivo}) & mL/min/kg & Dog & 1181 & 0.1 & 2946 & 12.6 & 15.2 \\ \hline
Plasma Clearance (\textit{In Vivo}) & mL/min/kg & Rat & 10431 & 0.001 & 8763 & 30.2 & 38.2 \\ \hline
Vd,ss & L/kg & Dog & 1054 & 0.07 & 569 & 1.9 & 1.9 \\ \hline
Vd,ss & L/kg & Rat & 9681 & 0.01 & 2080 & 2.3 & 2.4 \\ \hline
Hepatocyte Clearance & mL/min/g liver tissue & Human & 1695 & 0.01 & 97 & 1.6 & 1.5 \\ \hline
Hepatocyte Clearance & mL/min/g liver tissue & Dog & 630 & 0.1 & 504 & 2 & 1.8 \\ \hline
Hepatocyte Clearance & mL/min/g liver tissue & Rat & 2098 & 0.02 & 878 & 2.9 & 2.9 \\ \hline
Microsomal Clearance & mL/min/g liver tissue & Human & 29162 & 0 & 156 & 2.8 & 2.4 \\ \hline
Microsomal Clearance & mL/min/g liver tissue & Dog & 2080 & 0.03 & 150 & 2.5 & 1.8 \\ \hline
Microsomal Clearance & mL/min/g liver tissue & Rat & 30563 & 0.01 & 198 & 3.9 & 3.7 \\ \hline
LogD &  &  & 27345 & 0.01 & 53703 & 258 & 407 \\ \hline
\end{tabular}
\label{tab:PharmSum}
}
\end{table*}

\begin{table*}[]
\caption{Safety datasets used to benchmark \pipeline}

\resizebox{\textwidth}{!}{%
\begin{tabular}{|l|l|l|l|l|}
\hline
\textbf{Assay} & \textbf{Target} & \textbf{Primary Liability} & \textbf{Experimental System} & \textbf{Detection} \\ \hline
BSEP pIC50 & Bile Salt Export Pump & Hepatic & membrane vesicles &  \\ \hline
ADRA1B pIC50 & Adrenergic $\alpha$1B pIC50 & CNS & Intracell Ca &  \\ \hline
ADRA2C pIC50 & $\alpha$2C Adrenoceptor & CNS & CHO K1 &  \\ \hline
ADRB2 pEC50 & $\beta$2 Adrenoceptor & CNS & FRET &  \\ \hline
CHRM1 pEC50 & Cholinergic Receptor Muscarinic 1 & CNS & CHO & Intracellular Ca Fluorescence \\ \hline
CHRM1 pIC50 & Cholinergic Receptor Muscarinic 1 & CNS & CHO & Intracellular Ca Fluorescence \\ \hline
CHRM2 pEC50 & Cholinergic Receptor Muscarinic 2 & CNS & CHO & Intracellular Ca Fluorescence \\ \hline
DRD2 pEC50 & Dopamine D2 & CNS & HEK293F Low Na GTPgS & SPA \\ \hline
GRIN1 pIC50 & GRIN1 GRIN2B NR2B NR1A 2B Subunit pIC50 & CNS &  &  \\ \hline
HRH1 pIC50 & Histamine Receptor H1 & CNS &  & Luminescence \\ \hline
HTR1B pIC50 & 5-hydroxytryptamine receptor 1B & CNS & 10ul LEADseeker GTPgS &  \\ \hline
HTR2A pEC50 & 5-hydroxytryptamine Receptor 2A & CNS & HEK & Luminescence \\ \hline
HTR2A pIC50 & 5-hydroxytryptamine Receptor 2A & CNS & HEK & Luminescence \\ \hline
HTR2C  pEC50 & 5-hydroxytryptamine Receptor 2C & CNS & CHO & Luminescence \\ \hline
HTR2C) pIC50 & 5-hydroxytryptamine Receptor 2C & CNS & CHO & Luminescence \\ \hline
HTR3A pIC50 & 5-hydroxytryptamine Receptor 3A & CNS & FLIPR &  \\ \hline
KCNA5 (Kv1.5) pIC50 & KCNA5 (Kv1.5) & Cardiovascular & CHO & Electrophys \\ \hline
KCNE1 KCNQ1 (Kv7.1)  pIC50 & KCNE1 KCNQ1 (Kv7.1) & Cardiovascular & MinK Human Blocker CHO & Electrophys \\ \hline
MAOA pIC50 & Monoamine Oxidase A & CNS &  & FLINT \\ \hline
PDE3A pIC50 & Phosphodiesterase 3A & Cardiac &  & SPA(cAMP Inhibition) \\ \hline
PDE4B pIC50 & Phosphodiesterase  4B & CNS &  & SPA \\ \hline
Phospholipidosid pEC50 & Phospholipidosis Induction & Cellular Tox & HEPG2 & FLINT \\ \hline
PI3K$\gamma$ pIC50 & Phosphoinositide 3-kinase $\gamma$ (pI3K$\gamma$) & Cellular Tox &  & TR FRET \\ \hline
COX 2 pIC50 & Cyclooxygenase 2 & Cardiovascular &  & FLINT SAR \\ \hline
SCN5A (NaV1.5) pIC50 & SCN5A (NaV1.5) & Cardiovascular &  &  \\ \hline
SCL6A2 pIC50 & Noradrenaline Transporter NET & CNS &  & BacMam Bind SPA \\ \hline
SLC6A4 pIC50 & Seratonin Transporter (SERT) & CNS &  & BacMam binding SPA \\ \hline
OATP1B1 pIC50 & Organic Anion Transport Polypeptide (SLCO1B1) & Hepatic & HEK & Image \\ \hline
\end{tabular}%
\label{tab:safety}
}
\end{table*}

\subsection{Experimental design for regression pharmacokinetic models}
To evaluate AMPL's performance, we built a total of 11,552 models on 15 pharmacokinetic datasets and 26 bioactivity datasets. These models include 9,422 regression models and 2,130 classification models. 

We evaluated a variety of deep learning model types and architectures and compared them to baseline random forest models. We explored the performance of four types of features: ECFP fingerprints, MOE descriptors, Mordred descriptors, and graph convolution-based latent vectors. For the neural network models, we searched over many combinations of learning rates, numbers of layers, and nodes per layer. For each combination of neural network hyperparameters, we trained for up to 500 epochs and used a validation set performance metric ($R^2$ for regression, $ROC AUC$ for classification) to choose an early stopping epoch for the final model. For random forest models, the only hyperparameter varied was the maximum tree depth, as previous experiments showed that other model hyperparameters had a minimal effect for our datasets. The complete set of hyperparameters varied was as follows:
\begin{itemize}
    \item Splitter Types: scaffold and random
    \item Fraction for train set: 0.7
    \item Fraction for validation set: 0.1
    \item Fraction for holdout set: 0.2
    \item Feature types: ECFP, MOE, mordred, and graph convolution
    \item Model types: neural network and random forest
    \item Neural network learning rates: 0.0001, 0.00032, 0.001, 0.0032, 0.01 
    \item Maximum number of epochs: 500
    \item Number of layers: 1,2
    \item Layer size options: 1024,256,128, 64, 32,16,8,4,1
    \item Maximum final layer size: 16
    \item Dropout rate: 0.1
\end{itemize}

\subsection{Analysis of modeling performance}
\begin{figure}[htp]
    \centering
    \includegraphics[width=\columnwidth]{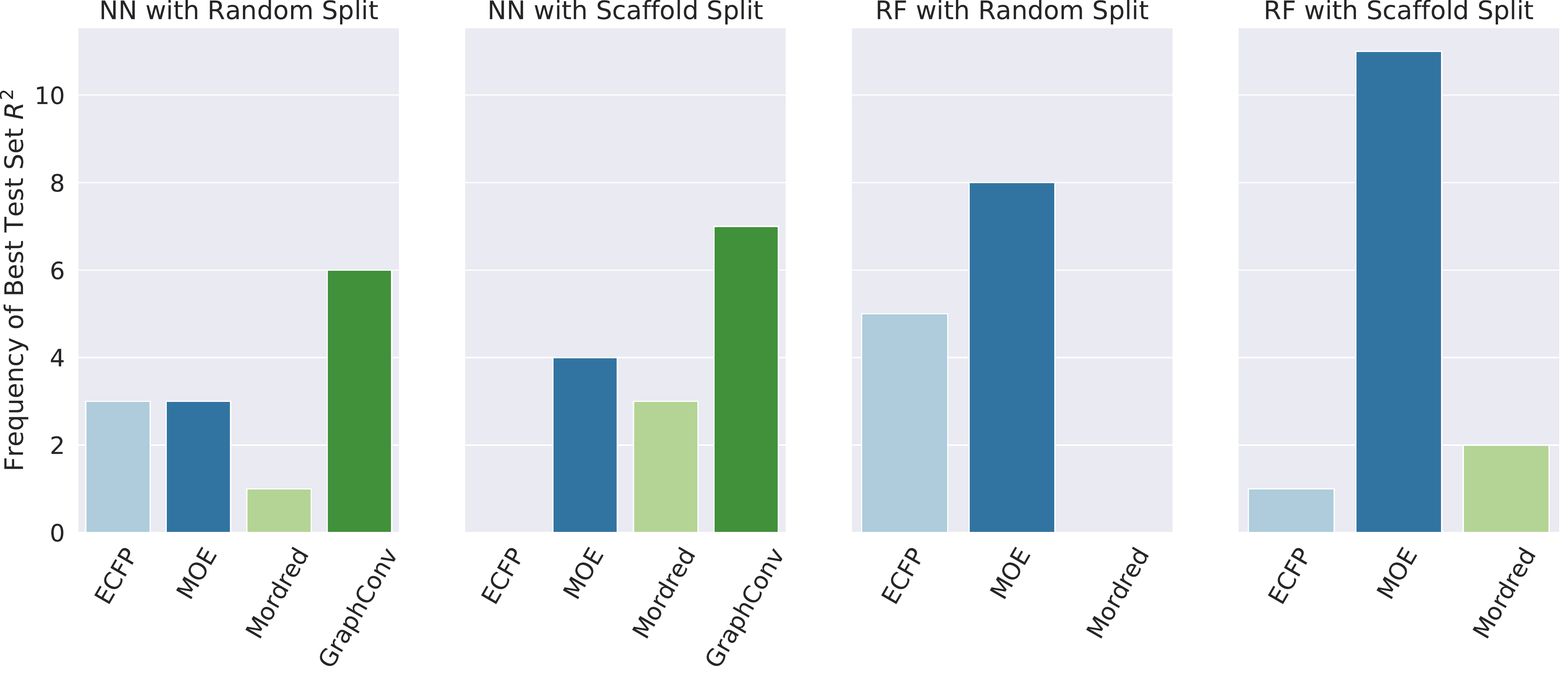}
    \caption{Number of times each featurization type produces the best model for the 15 PK datasets}
    \label{fig:regression_feat_perf}
\end{figure}

To identify which featurization type generated the most predictive models for each model type, models with the best validation set $R^2$ score were selected for each model/splitter/dataset combination. The number of ``best" models for which each feature type yielded the highest test set $R^2$ score is plotted in Figure \ref{fig:regression_feat_perf}. 
Figure \ref{fig:regression_feat_perf} shows that the chemical descriptors generated by the commercial MOE software outperformed those produced by the open source Mordred package in most cases. DeepChem's graph convolution networks outperform all other feature types for neural network models.

\begin{figure}[h!]
    \centering
    \includegraphics[width=\columnwidth]{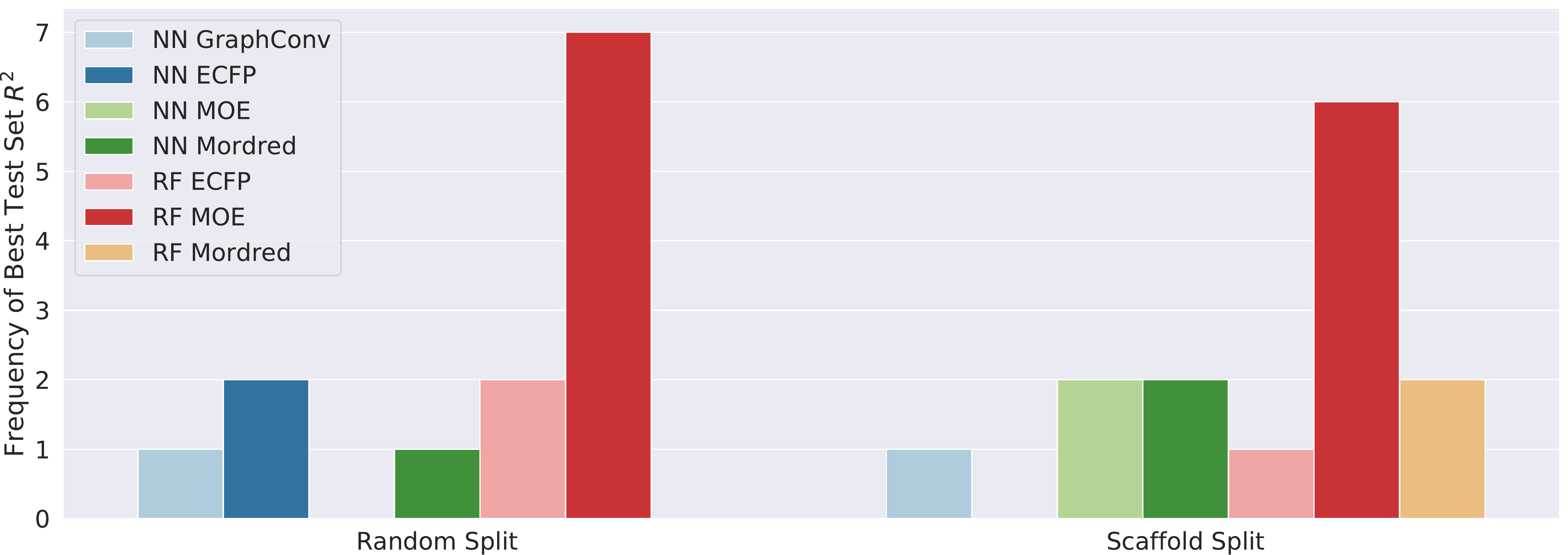}
    \caption{Number of times each featurization type/ model type combination produces the best model for the 15 PK datasets}
    \label{fig:regression_test_model_feat_perf}
\end{figure}

The model/featurization combination with the most accurate predictions on the holdout set is shown in  Figure \ref{fig:regression_test_model_feat_perf}. MOE featurization with random forest models most frequently outperformed other featurization/model type combinations for both types of splitters.

\begin{figure}[htp]
    \centering
    \includegraphics[width=\columnwidth]{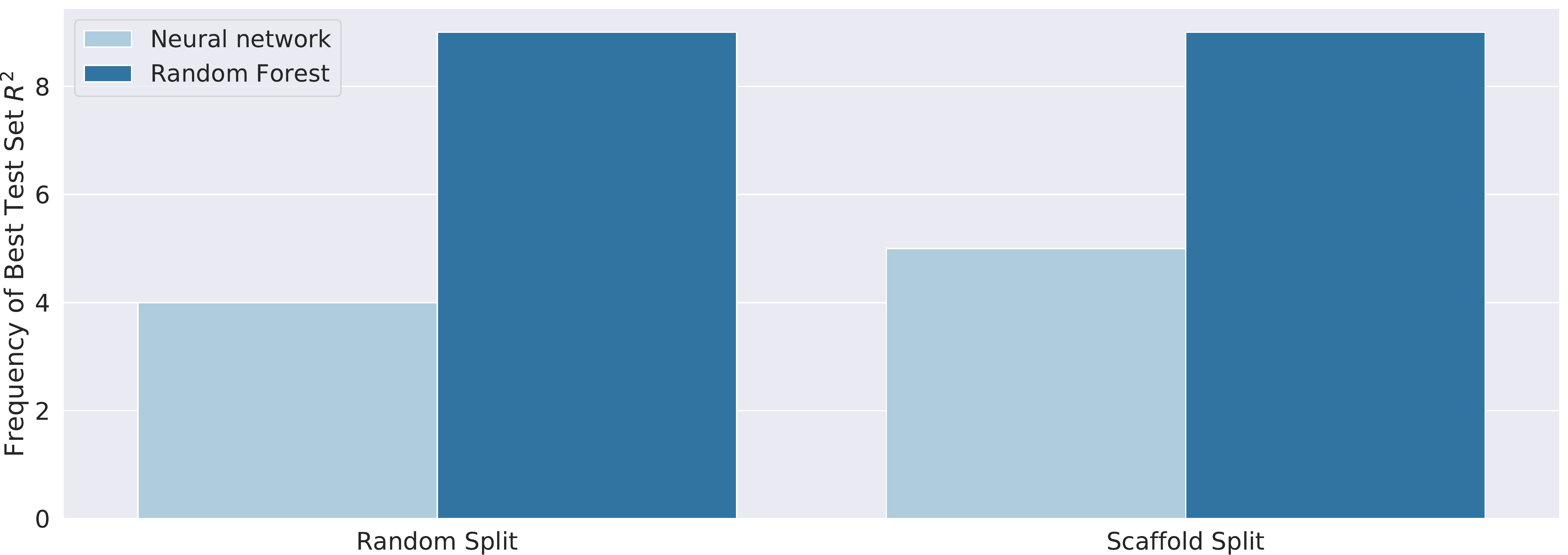}
    \caption{Number of times each model type produces the best model for the 15 PK datasets}
    \label{fig:regression_model_perf}
\end{figure}

Figure \ref{fig:regression_model_perf} confirms that random forest models tend to outperform neural network models for the evaluated datasets. 

\subsection{Investigation into neural network performance}

\begin{figure}[htp]
    \centering
    \includegraphics[width=\columnwidth]{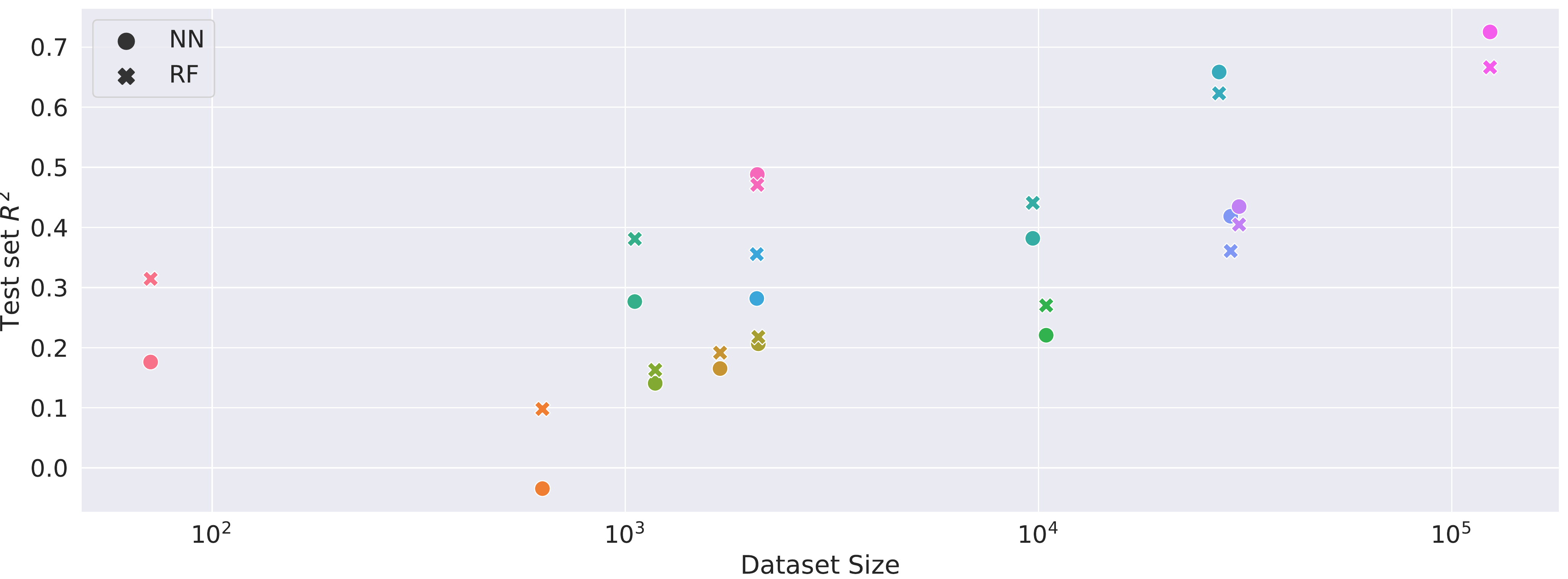}
    \caption{Plot of best test set $R^2$ values versus the dataset size for neural network and random forest models}
    \label{fig:regression_num_samples_perf}
\end{figure}

\begin{figure}[htp]
    \centering
    \includegraphics[width=\columnwidth]{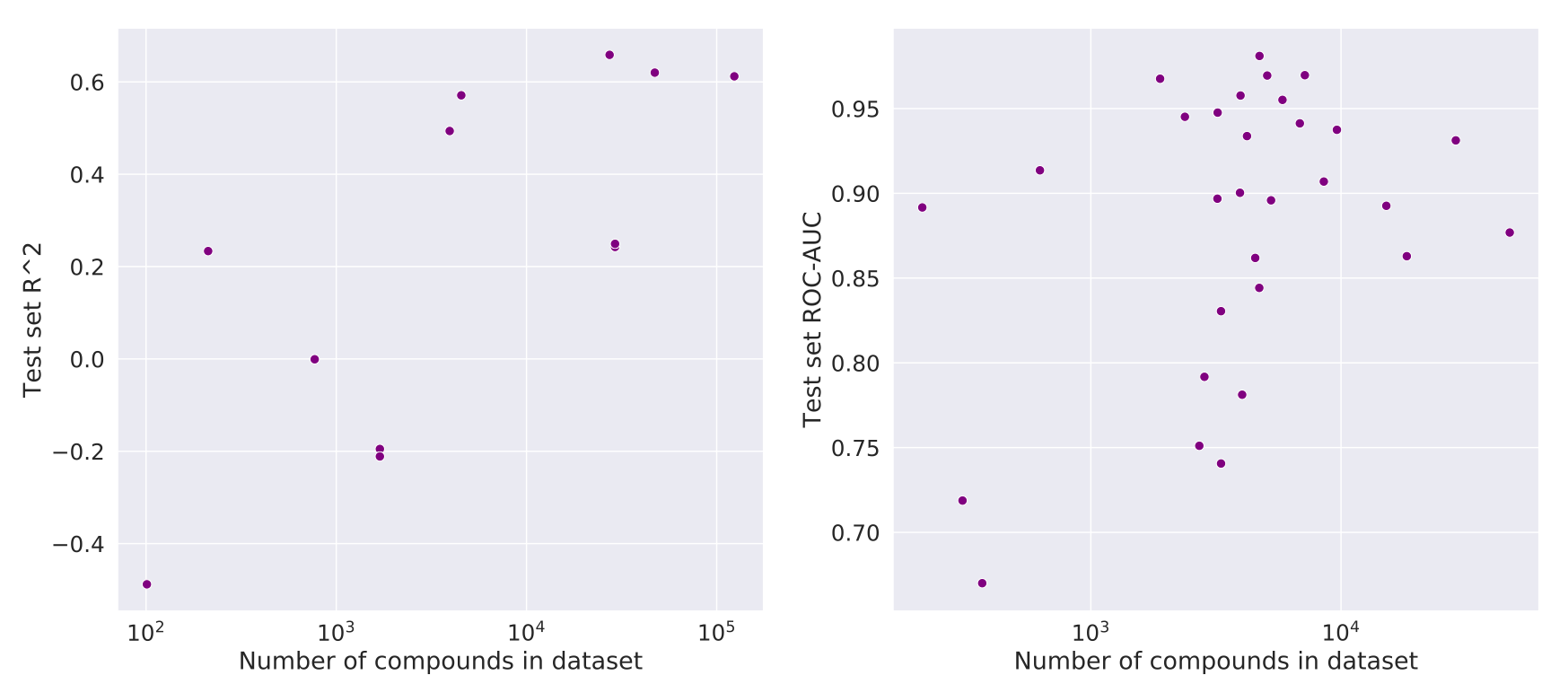}
    \caption{Per-dataset model accuracy versus dataset size}
    \label{fig:numcpds_perf}
\end{figure}

Neural networks are known to perform more poorly on smaller datasets, so we wanted to examine the relationship between the size of a dataset and the test set $R^2$ values for the best random forest and neural network models for that dataset. Figure \ref{fig:regression_num_samples_perf} shows the test set $R^{2}$ values for the best neural network and random forest models for each dataset, where best is defined as the model with the highest validation set $R^2$ value. The figure shows that as the dataset size increases, the $R^2$ score for the test set increases as well.
 The pattern is true for the overall best model, regardless of type, for both regression and classification, as shown in Figure \ref{fig:numcpds_perf}. These results indicate that we will need to augment our datasets to further improve model performance. We plan to explore multiple avenues to address this requirement: conducting additional experiments, running simulations, sourcing public data, building multi-task models, and experimenting with transfer learning approaches. 

\begin{figure}[htp]
    \centering
    \includegraphics[width=\columnwidth]{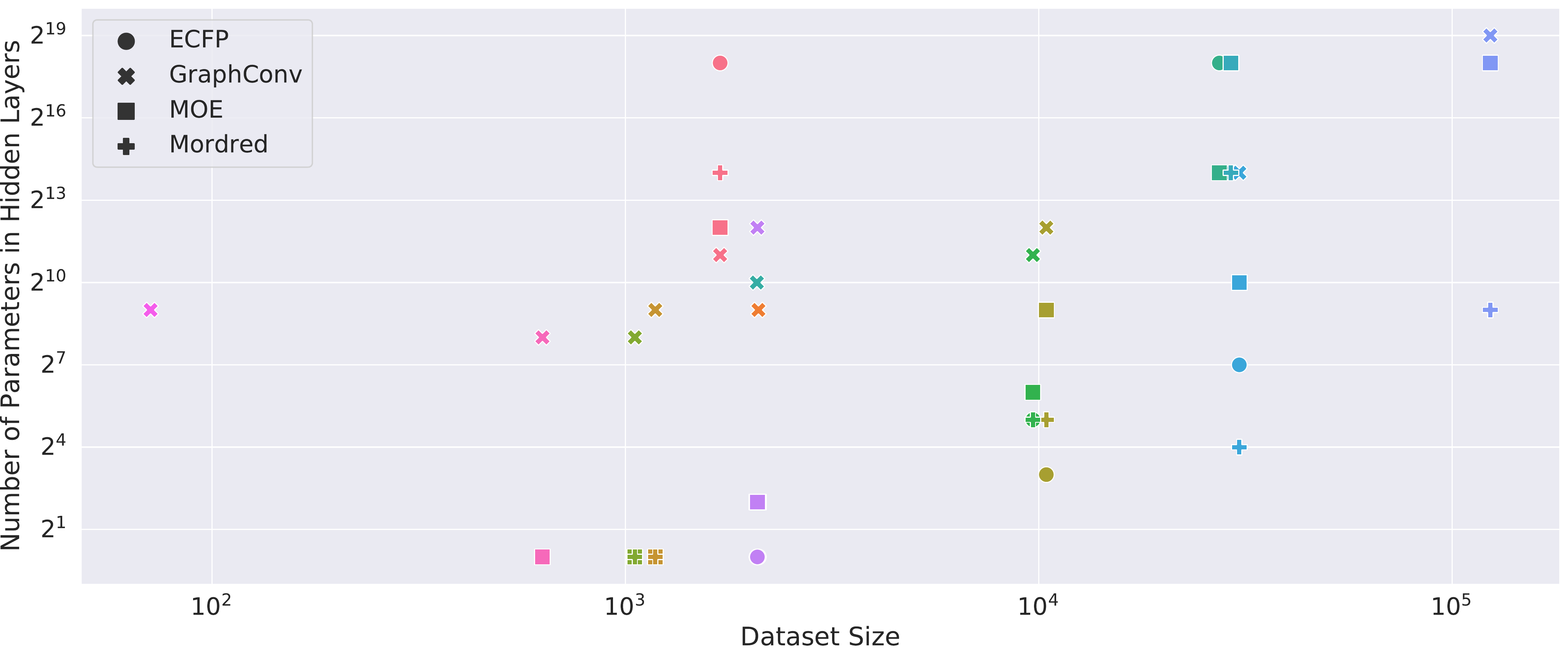}
    \caption{Number of hidden layer parameters versus number of samples for the best model for each dataset/featurizer combination}
    \label{fig:params_samples}
\end{figure}

We also examined the architectures that yielded the best model for each feature type for the neural network models. Our hypothesis was that larger datasets would perform better with larger networks. Figure \ref{fig:params_samples} shows number of parameters in the hidden layers of the model versus the size of the dataset. The color indicates the dataset and the shape indicates the featurizer type. The number of parameters for the 2-layer networks was calculated by multiplying the first and second layers together. We can see a clear lower bound in the number of parameters for the best network for all featurizer types as the dataset size increases.

\subsection{Summary of model performance}
\begin{figure}[htp]
    \centering
    \includegraphics[width=\columnwidth]{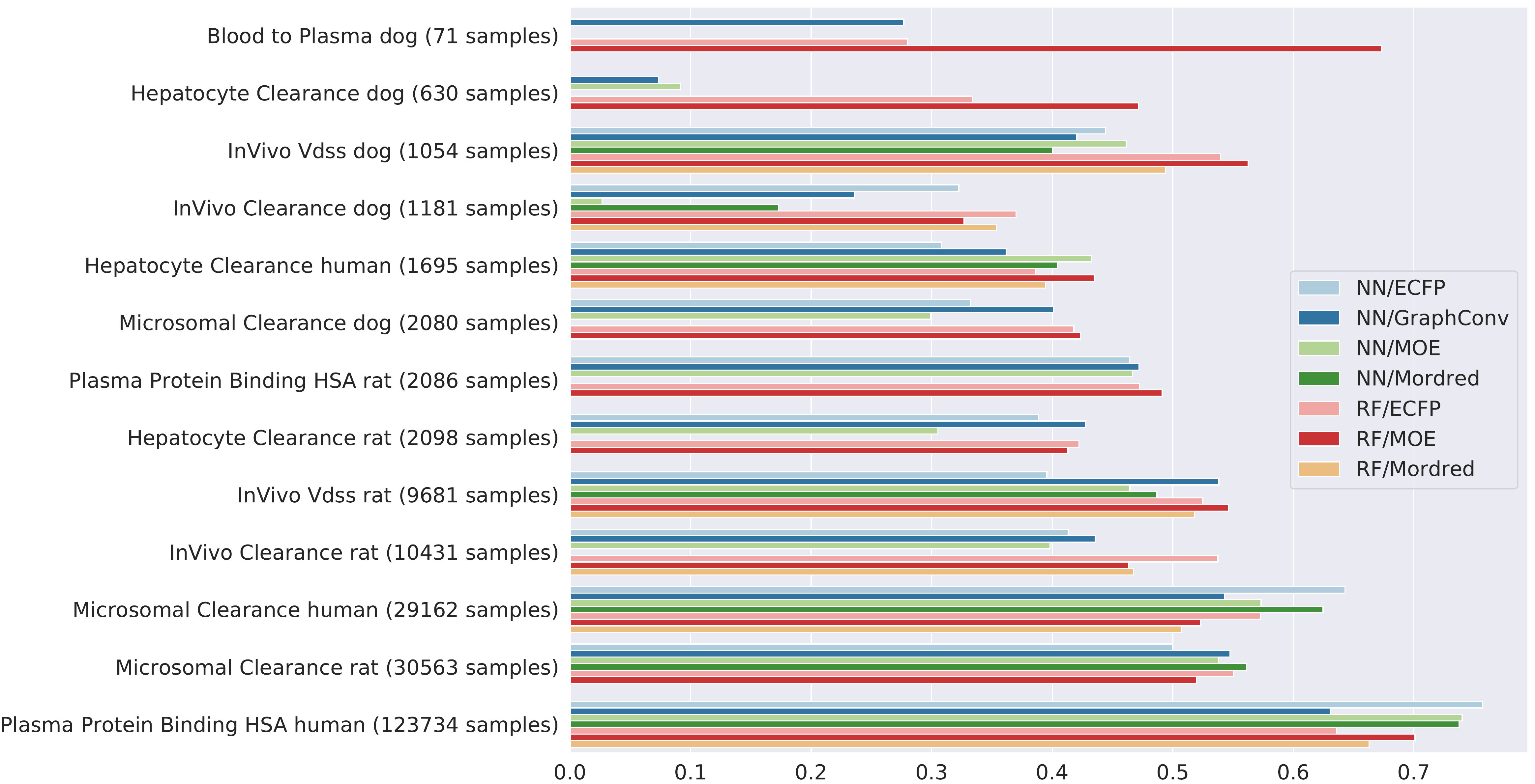}
    \caption{Performance accuracy for regression for random split}
    \label{fig:regression_perf_random}
\end{figure}

\begin{figure}[htp]
    \centering
    \includegraphics[width=\columnwidth]{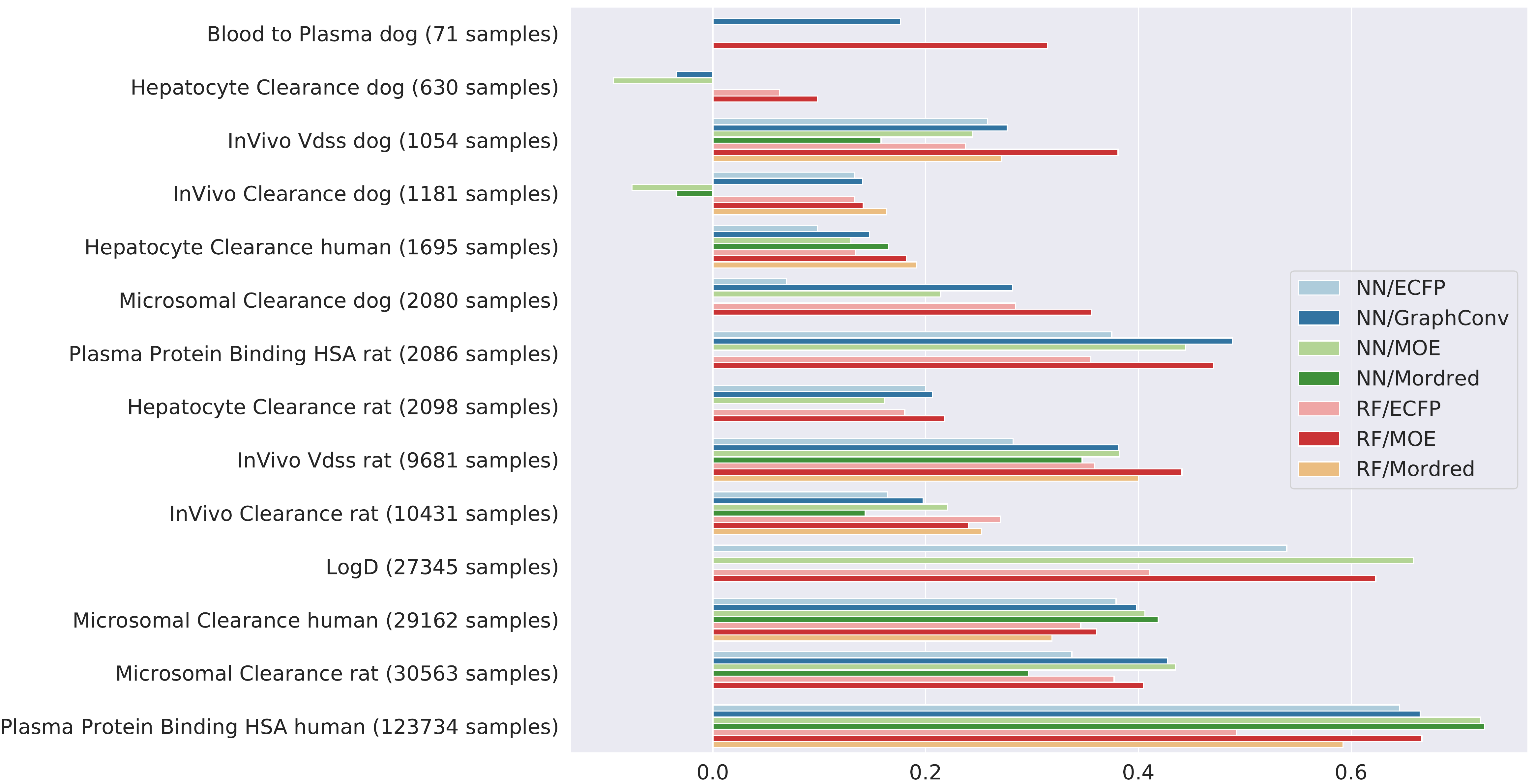}
    \caption{Performance accuracy for regression for scaffold split}
    \label{fig:regression_perf_scaffold}
\end{figure}

Figure \ref{fig:regression_perf_random} and Figure \ref{fig:regression_perf_scaffold} show the full set of test set $R^2$ values for the best model for each molecular featurization representation and model type for random and scaffold splits respectively (picked as before by the best validation set $R^2$ value). Random sampling inflates the $R^2$ values of the holdout set, which is as expected since there is greater structural overlap between the set of compounds in the training and holdout set.
For scaffold split-generated holdout sets, there is a very clear pattern between dataset size and $R^2$ value, although the complexity of the predicted property and quality of the dataset also obviously has an effect.

\subsection{Model tuning results}
To evaluate whether hyperparameter search improves model performance, the test set $R^2$ for a baseline model was compared with the test set $R^2$ from the best-performing model selected by looking at the validation set $R^2$ value. Small datasets and ECFP-based models, which showed poor neural network performance overall, showed little to no improvement, while better-performing datasets and featurizers showed greater improvement with hyperparameter search. This suggests that data augmentation will be necessary to improve prediction performance on the smaller problematic datasets, and that ECFP is a poor featurizer no matter the hyperparameters.

\begin{figure}[htp]
    \centering
    \includegraphics[width=\columnwidth]{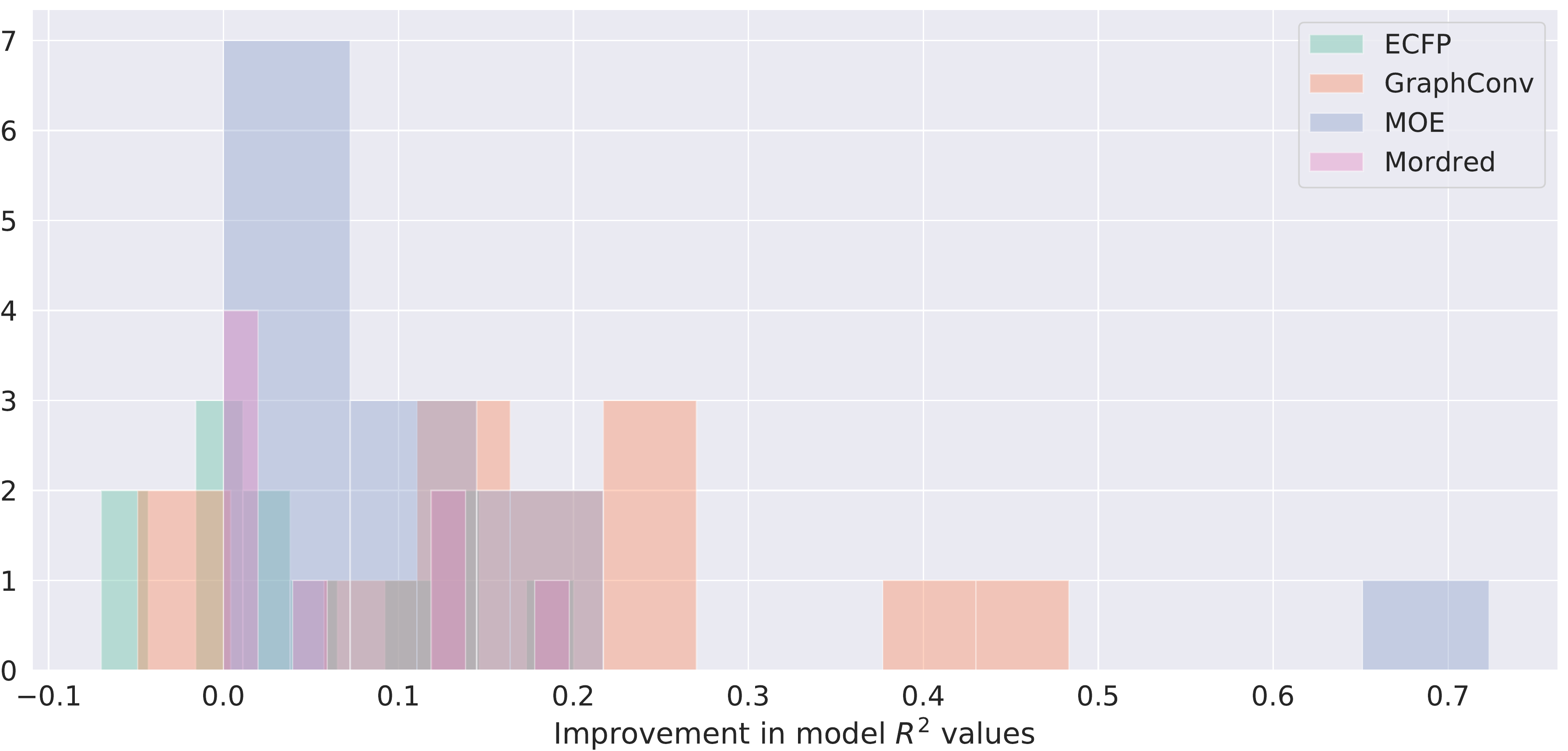}
    \caption{Histogram of improvement in $R^2$ values for the test set for the four featurizers for neural network models}
    \label{fig:hyperparam_perf}
\end{figure}

\subsection{Classification experiments}
A set of classification model experiments were also conducted for a panel of 28 bioactivity datasets, without any hyperparameter tuning. In total 2,130 neural network and random forest models were generated. A dose concentration threshold was used to label active and inactive compounds on a per-dataset basis using thresholds provided by domain experts at GlaxoSmithKline. The classes were extremely unbalanced, which partially explains the high ROC-AUC scores.
\begin{figure}[htp]
    \centering
    \includegraphics[width=\columnwidth]{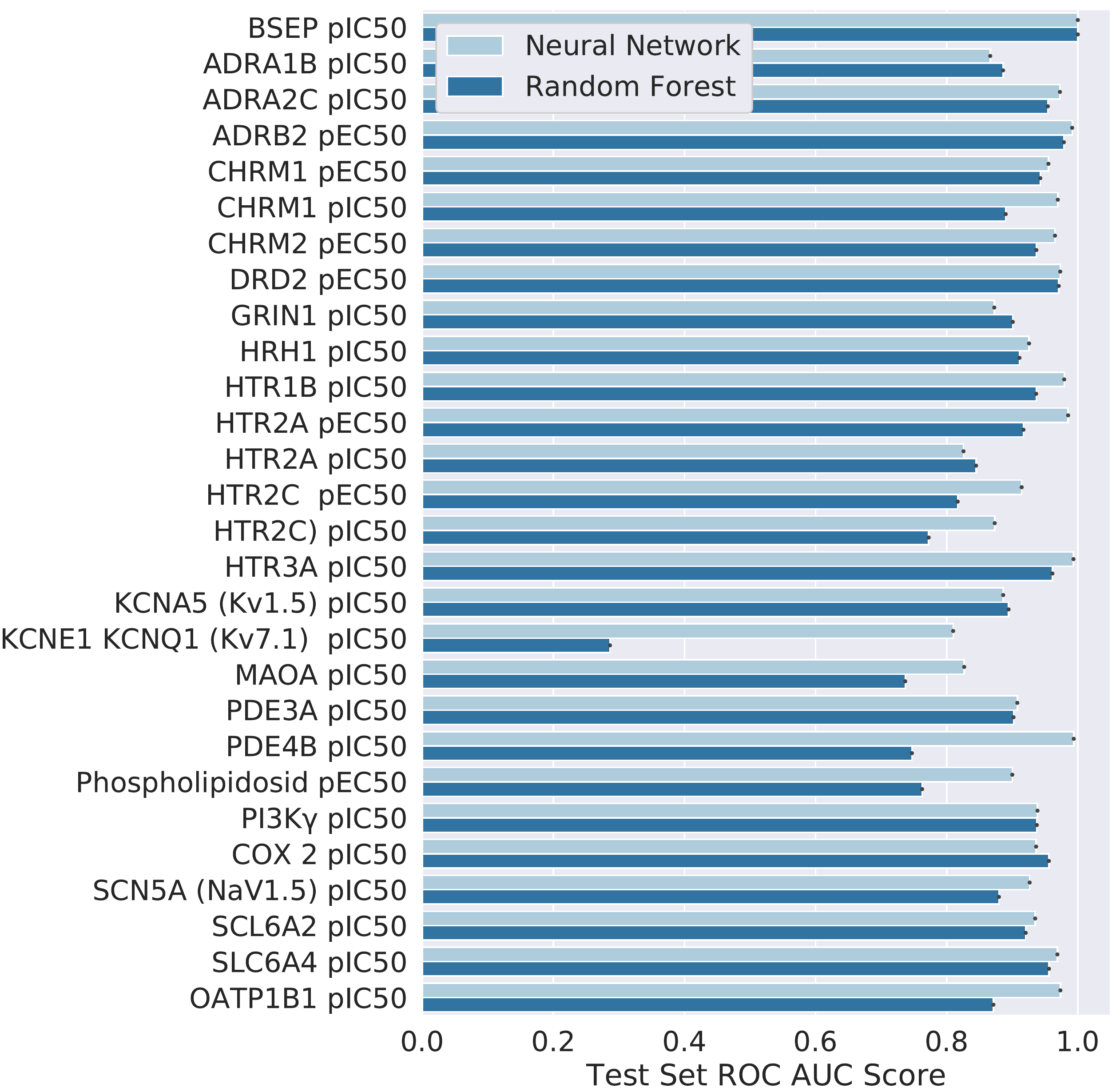}
    \caption{Performance accuracy for classification}
    \label{fig:class_perf}
\end{figure}

\subsection{Uncertainty quantification}
To explore the utility of the uncertainty quantification values produced by neural network and random forest models, a case study is presented for three representative PK parameter datasets: rat plasma clearance (\textit{in vivo}), human microsomal clearance, and human plasma protein binding HSA. These datasets were selected to represent small, medium, and large sized datasets with low, medium, and high $R^2$ values. 

\subsubsection{Precision-recall plot analysis}
Precision-recall curves measure the fraction of low error predictions made at varying UQ thresholds. Precision is defined as the fraction of predictions with UQ values less than the UQ threshold, with error less than some predefined threshold. For this analysis we use mean logged error and define ``low-error" as samples with logged error below the mean (log served to normalize the distribution). Recall reports the fraction of low-error samples which pass the UQ filter threshold. Overall, we would like to use the UQ value as a threshold to identify low error samples at a higher rate than in the overall test set. Table \ref{tab:errors} shows the percentage of low error samples in the test set as a whole for each dataset/model/featurizer combination.

\begin{table}[]
\resizebox{\columnwidth}{!}{%
\begin{tabular}{|l|l|c|}
\hline
\textbf{Dataset} & \textbf{Model and featurizer type} & \textbf{\makecell{Percent of total \\low error samples}} \\ \hline
Rat Plasma Clearance (\textit{In Vivo}) & Neural network + ECFP & 41.4\% \\ \hline
Rat Plasma Clearance (\textit{In Vivo}) & Neural network + GraphConv & 41.8\% \\ \hline
Rat Plasma Clearance (\textit{In Vivo}) & Neural network + MOE & 42.9\% \\ \hline
Rat Plasma Clearance (\textit{In Vivo}) & Neural network + Mordred & 40.5\% \\ \hline
Rat Plasma Clearance (\textit{In Vivo}) & Random forest + ECFP & 42.5\% \\ \hline
Rat Plasma Clearance (\textit{In Vivo}) & Random forest + MOE & 41.7\% \\ \hline
Rat Plasma Clearance (\textit{In Vivo}) & Random forest + Mordred & 42.0\% \\ \hline
Human Microsomal Clearance & Neural network + ECFP & 41.0\% \\ \hline
Human Microsomal Clearance & Neural network + GraphConv & 41.0\% \\ \hline
Human Microsomal Clearance & Neural network + MOE & 39.0\% \\ \hline
Human Microsomal Clearance & Neural network + Mordred & 39.8\% \\ \hline
Human Microsomal Clearance & Random forest + ECFP & 39.5\% \\ \hline
Human Microsomal Clearance & Random forest + MOE & 38.5\% \\ \hline
Human Microsomal Clearance & Random forest + Mordred & 39.6\% \\ \hline
Human Plasma Protein Binding HSA & Neural network + ECFP & 43.4\% \\ \hline
Human Plasma Protein Binding HSA & Neural network + GraphConv & 43.0\% \\ \hline
Human Plasma Protein Binding HSA & Neural network + MOE & 43.1\% \\ \hline
Human Plasma Protein Binding HSA & Neural network + Mordred & 43.5\% \\ \hline
Human Plasma Protein Binding HSA & Random forest + ECFP & 42.0\% \\ \hline
Human Plasma Protein Binding HSA & Random forest + MOE & 42.8\% \\ \hline
Human Plasma Protein Binding HSA & Random forest + Mordred & 42.5\% \\ \hline
\end{tabular}%
\caption{Percent of total low-error samples in the test set for the specified dataset, model/featurizer combinations}
\label{tab:errors}
}
\end{table}

\begin{figure}[htp]
    \centering
    \includegraphics[width=\columnwidth]{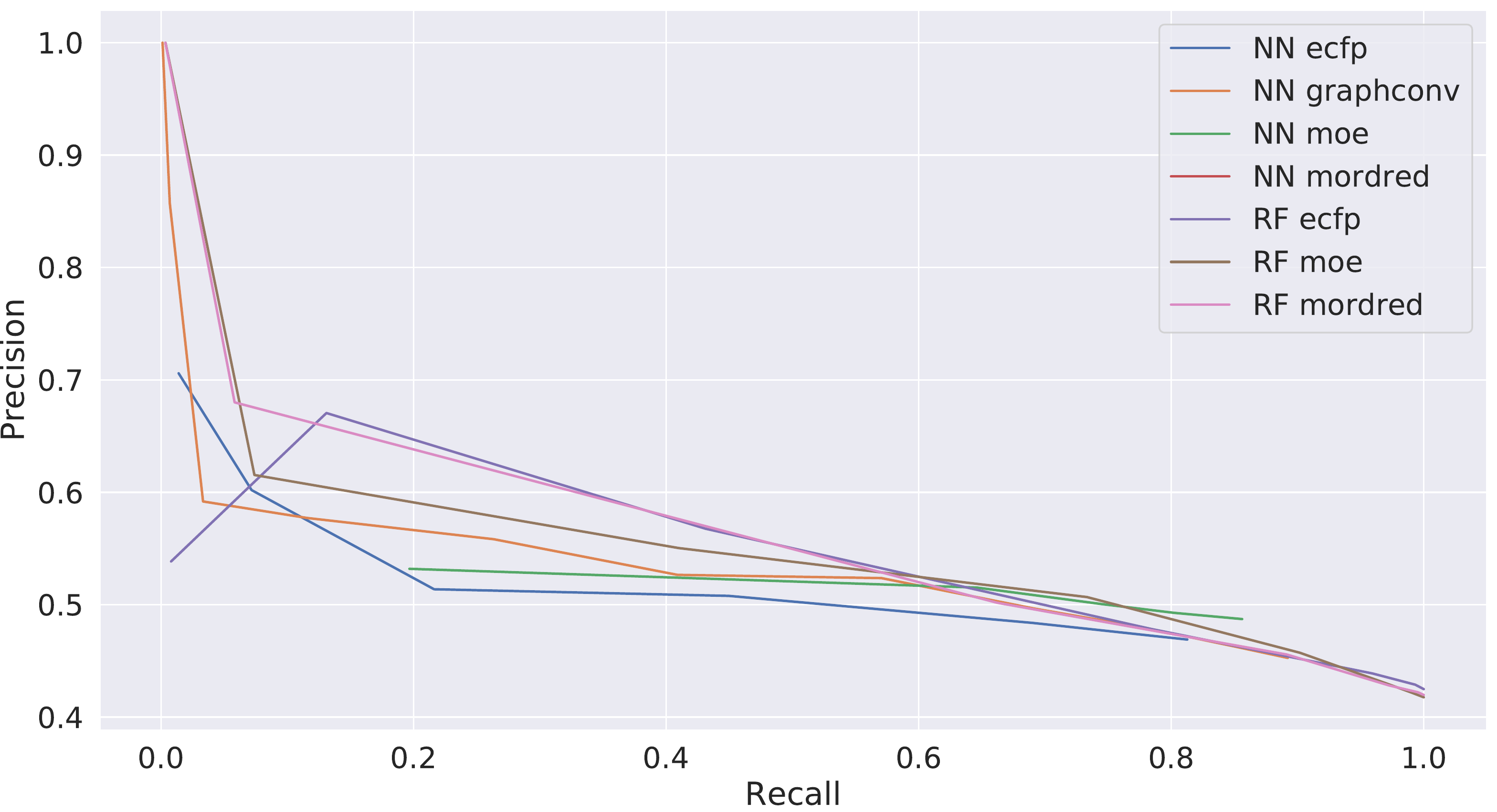}
    \caption{Precision-recall plot for rat plasma clearance (\textit{in vivo}), varying UQ value}
    \label{fig:invivo_rat_pr_uq}
\end{figure}

\begin{figure}[htp]
    \centering
    \includegraphics[width=\columnwidth]{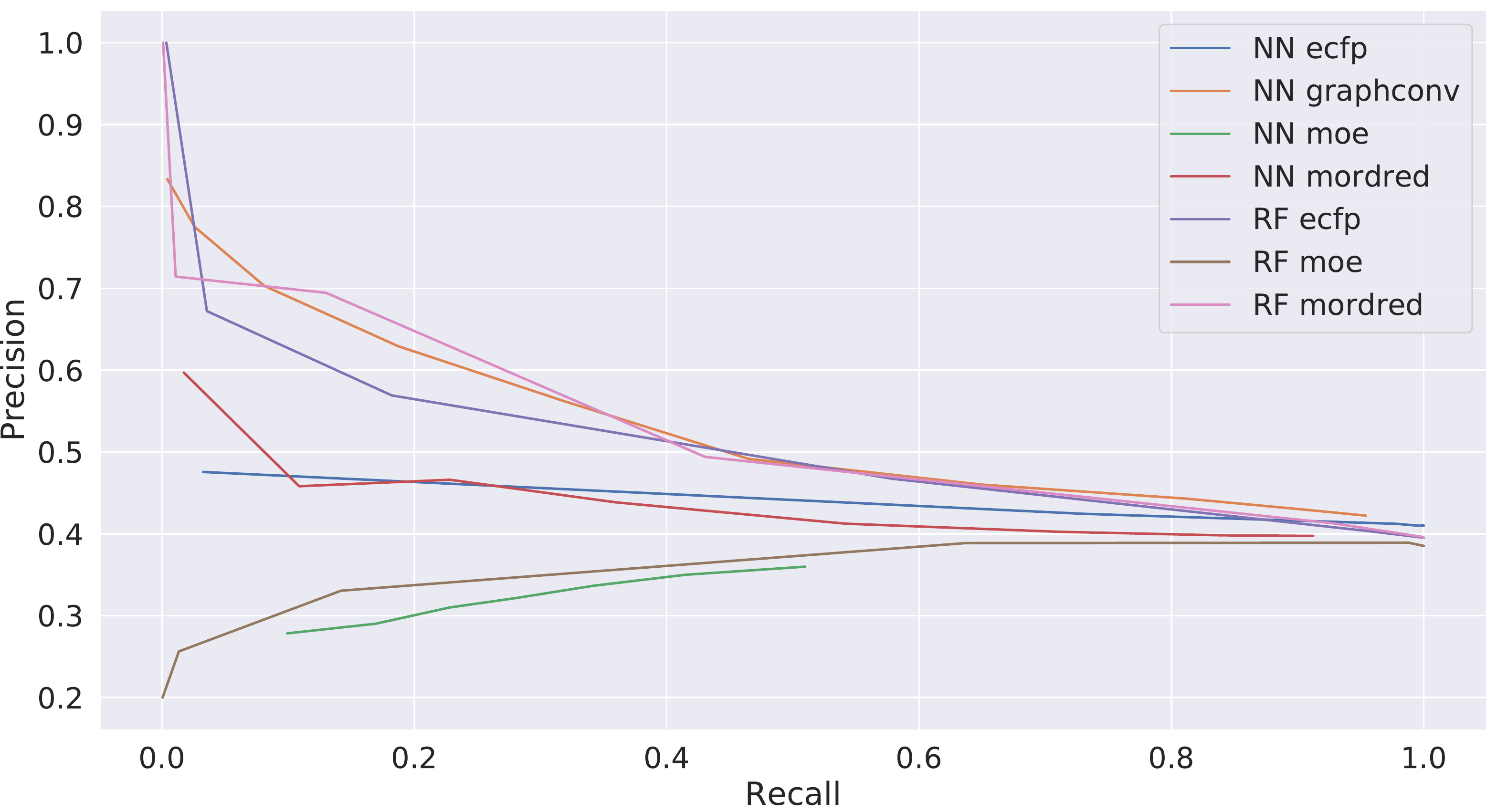}
    \caption{Precision-recall plot for human microsomal clearance, varying UQ value}
    \label{fig:microsomal_human_pr_uq}
\end{figure}

\begin{figure}[htp]
    \centering
    \includegraphics[width=\columnwidth]{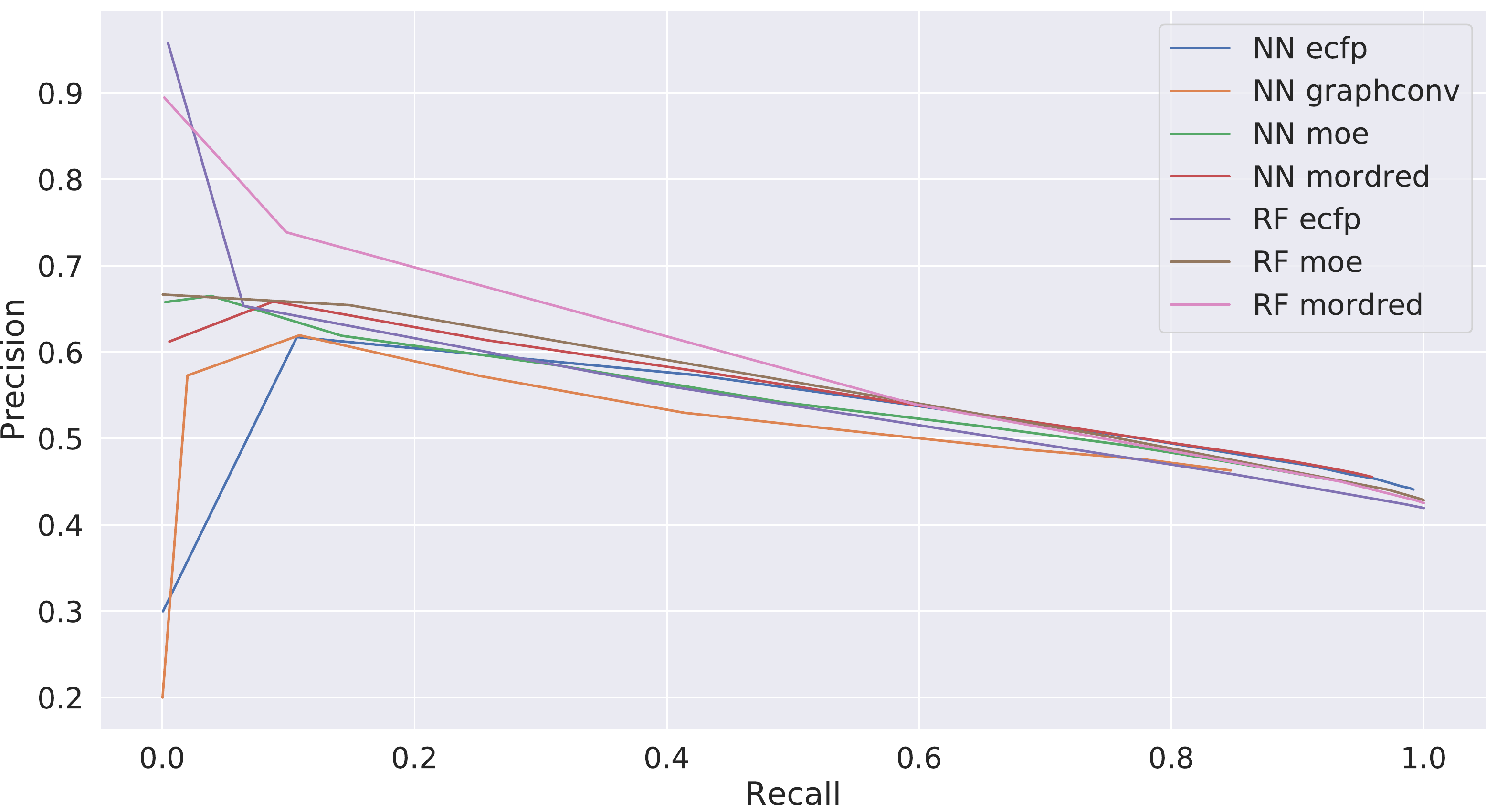}
    \caption{Precision-recall plot for human plasma protein binding HSA, varying UQ value}
    \label{fig:ppb_human_pr_uq}
\end{figure}

In general, a low UQ threshold with accurate uncertainty would correspond to a precision of 1, which means confident predictions correspond to low-error predictions. To have the greatest utility, the curve should keep fairly high precision as the recall increases. UQ successfully filters out low confidence predictions in some cases but performance varies widely with the model/featurization type and the dataset. Figures \ref{fig:invivo_rat_pr_uq}, \ref{fig:microsomal_human_pr_uq} and \ref{fig:ppb_human_pr_uq} show that precision drops quickly as recall increases and for some models precision is poor even when applying the lowest UQ threshold. Nevertheless, for each dataset there exists a UQ threshold for at least one model which could be used to increase the fraction of low error predictions over the baseline percentages shown in Table \ref{tab:errors}. For example, Figure \ref{fig:ppb_human_pr_uq} suggests that applying a UQ threshold could increase precision to 65\% from around 42\% with a recall of 10\%.  Later it is shown that for the human plasma protein binding HSA dataset, this could still yield a collection of compounds with a diverse range of response values.

\subsubsection{Calibration curves}
To further investigate how error changes as the uncertainty increases, we  plotted calibration curves of mean error per uncertainty bucket, with the 95\% confidence interval of error shown for each bucket as error bars. We would like uncertainty to serve as a proxy for error, so we would hope to see the mean error for the samples in a bucket to increase as the UQ threshold for that bucket increased. Results for neural network and random forest models built on MOE feature vectors and neural network graph convolution models are shown to demonstrate the variation in performance.

For rat plasma clearance (\textit{in vivo}), there is an overall upward trend for all three calibration curves, but it is not completely monotonically increasing for any of them. This is the smallest dataset of our case study, so increasing the bucket size may improve the choppiness of these curves, but overall UQ does not look like it would be a reliable proxy for error for this dataset.
\begin{figure}[htp]
    \centering
    \includegraphics[width=\columnwidth]{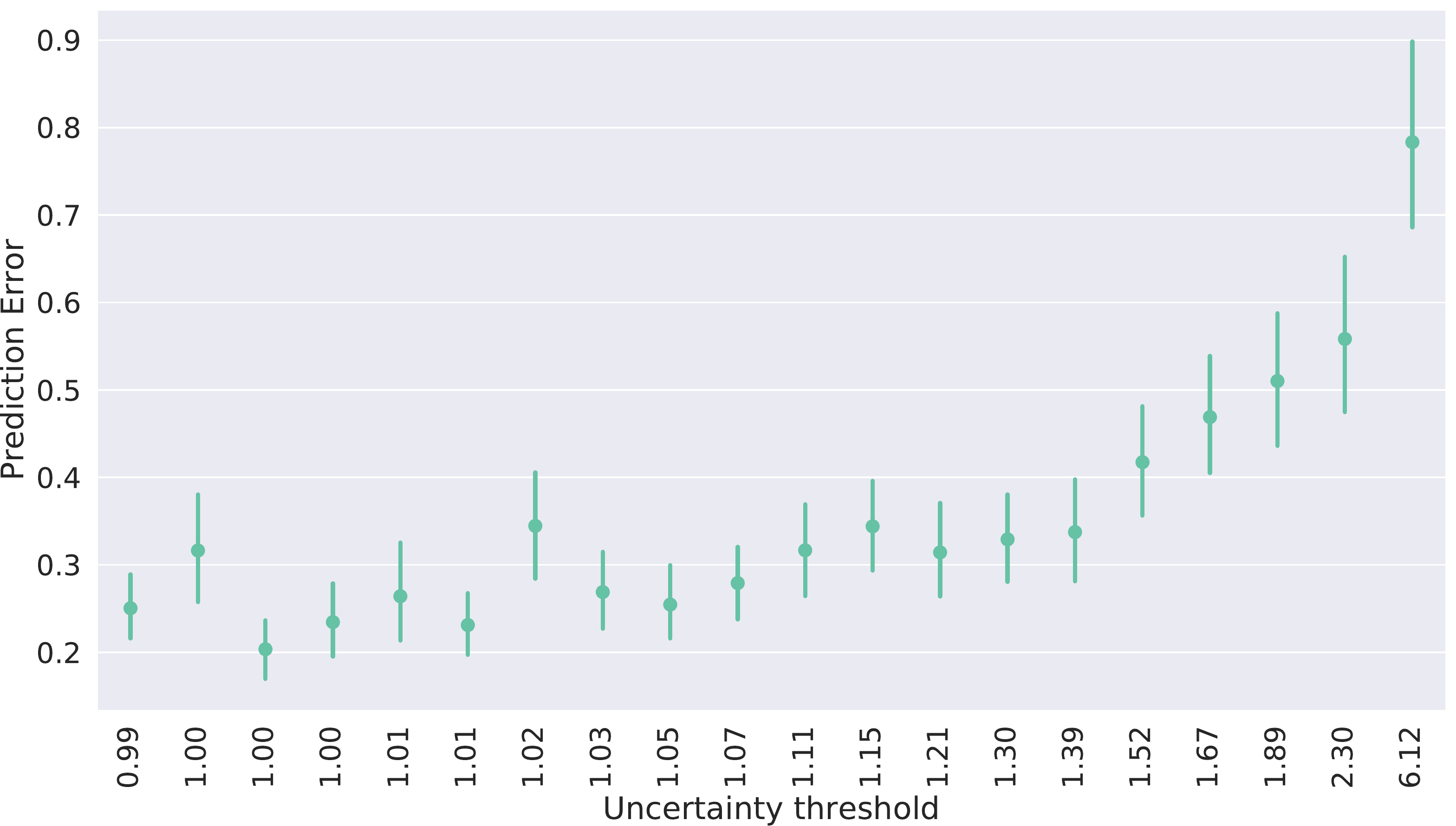}
    \caption{Mean error per uncertainty bucket for rat plasma clearance (\textit{in vivo}) neural network model with MOE features}
    \label{fig:invivo_nn_moe_ci}
\end{figure}

\begin{figure}[htp]
    \centering
    \includegraphics[width=\columnwidth]{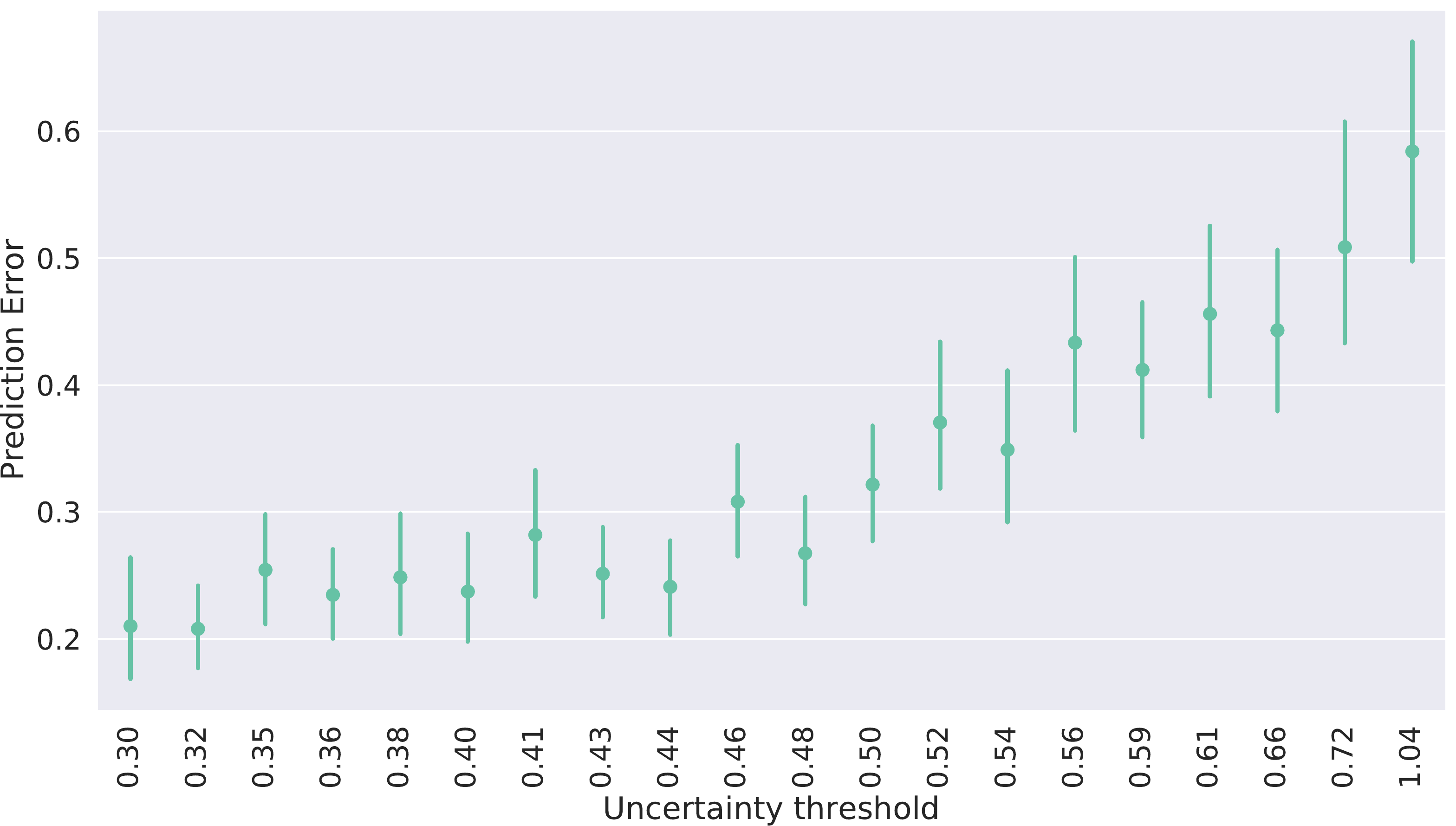}
    \caption{Mean error per uncertainty bucket for rat plasma clearance (\textit{in vivo}) random forest model with MOE features}
    \label{fig:invivo_rf_moe_ci}
\end{figure}

\begin{figure}[htp]
    \centering
    \includegraphics[width=\columnwidth]{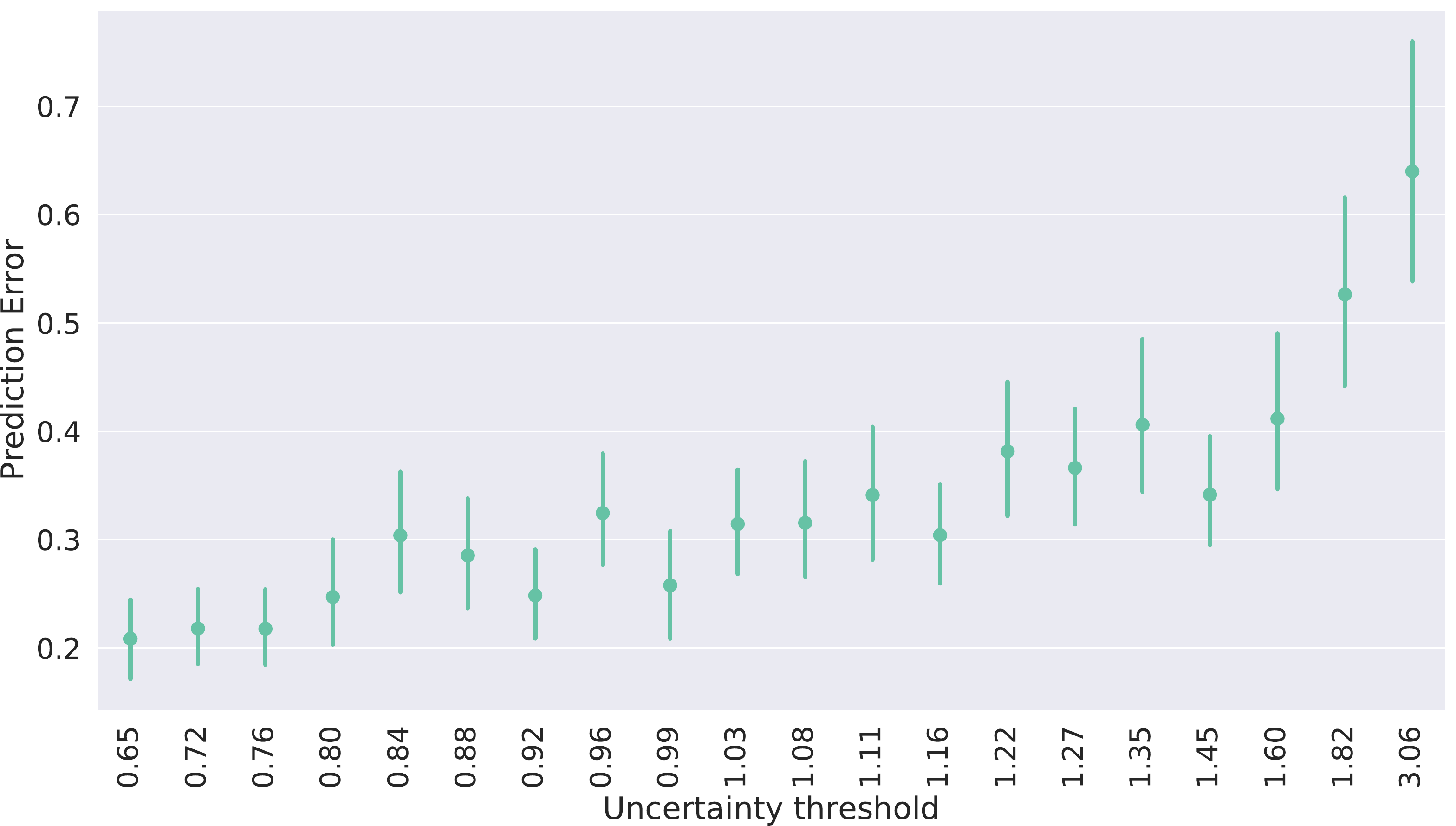}
    \caption{Mean error per uncertainty bucket for rat plasma clearance (\textit{in vivo}) neural network model with Graph Convolution features}
    \label{fig:invivo_nn_gc_ci}
\end{figure}

Human microsomal clearance shows greater variation in the calibration curves. For MOE features with a neural network model, shows an inverse pattern where the error actually decreases as the uncertainty increases.
\begin{figure}[htp]
    \centering
    \includegraphics[width=\columnwidth]{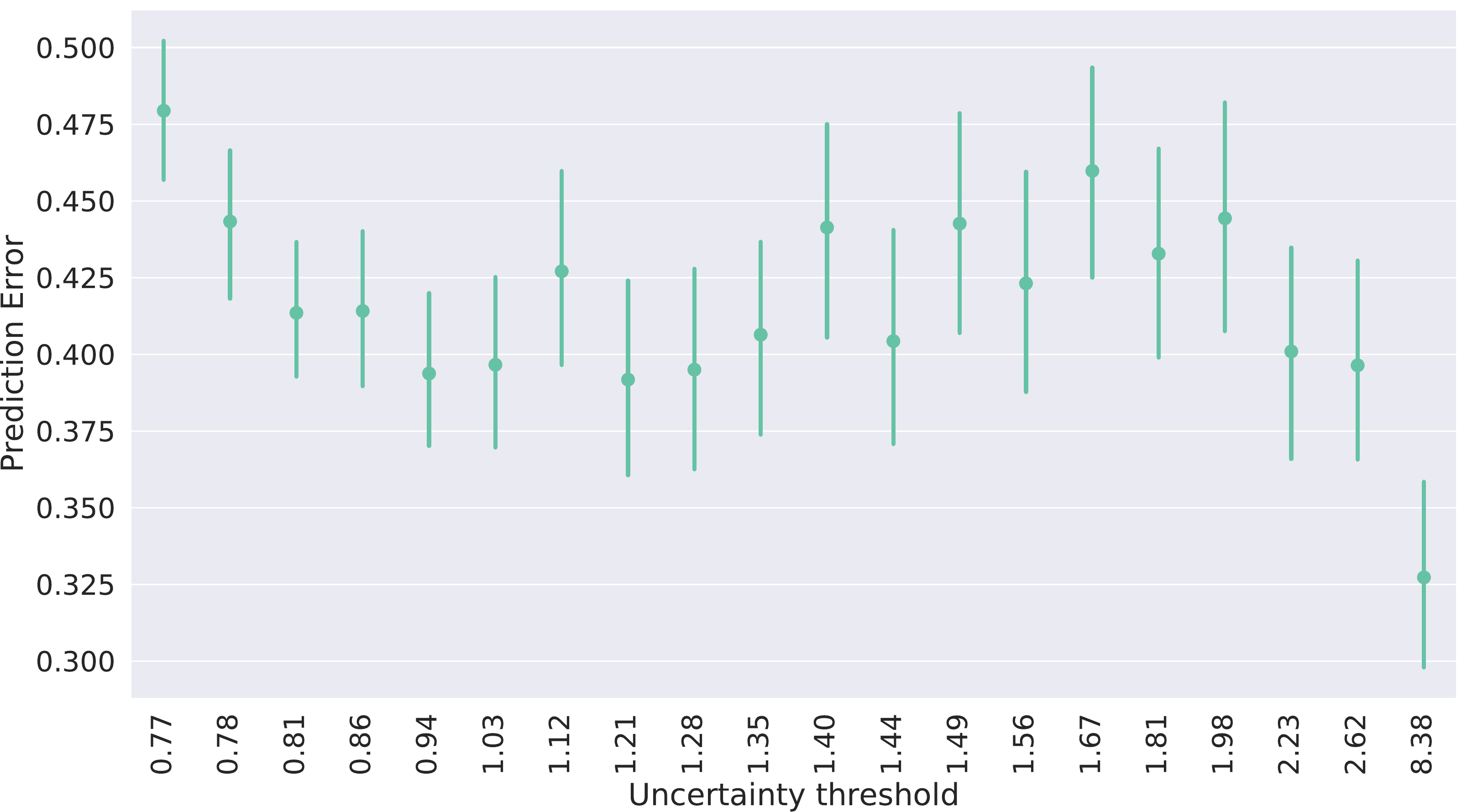}
    \caption{Mean error per uncertainty bucket for human microsomal clearance neural network model with MOE features}
    \label{fig:microsomal_nn_moe_ci}
\end{figure}
For MOE features with a random forest model, there seems to be no correlation, except for in the very highest bucket.
\begin{figure}[htp]
    \centering
    \includegraphics[width=\columnwidth]{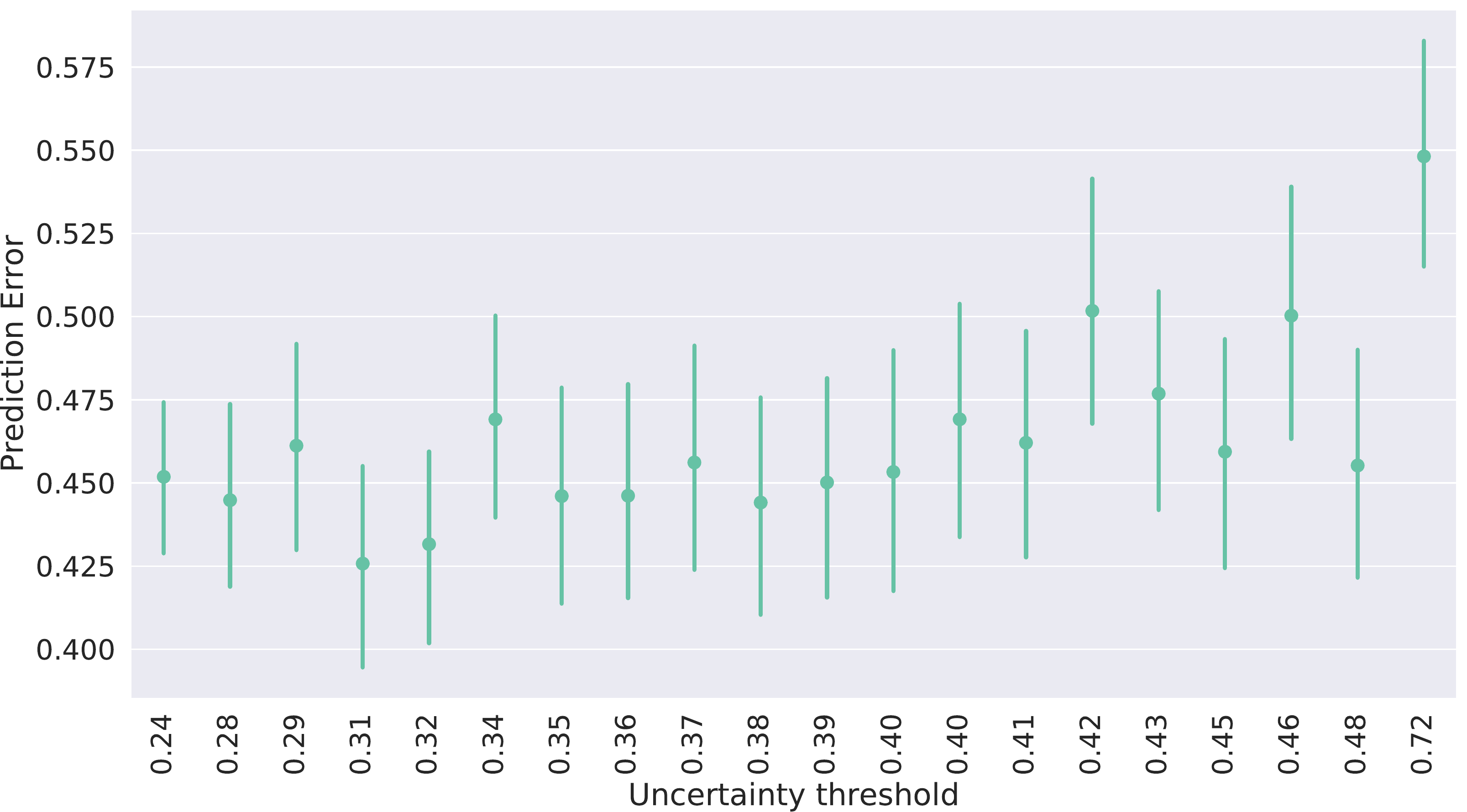}
    \caption{Mean error per uncertainty bucket for human microsomal clearance random forest model with MOE features}
    \label{fig:microsomal_rf_moe_ci}
\end{figure}
The graph convolution model, conversely, shows an upward trend, although it is not monotonically increasing.
\begin{figure}[htp]
    \centering
    \includegraphics[width=\columnwidth]{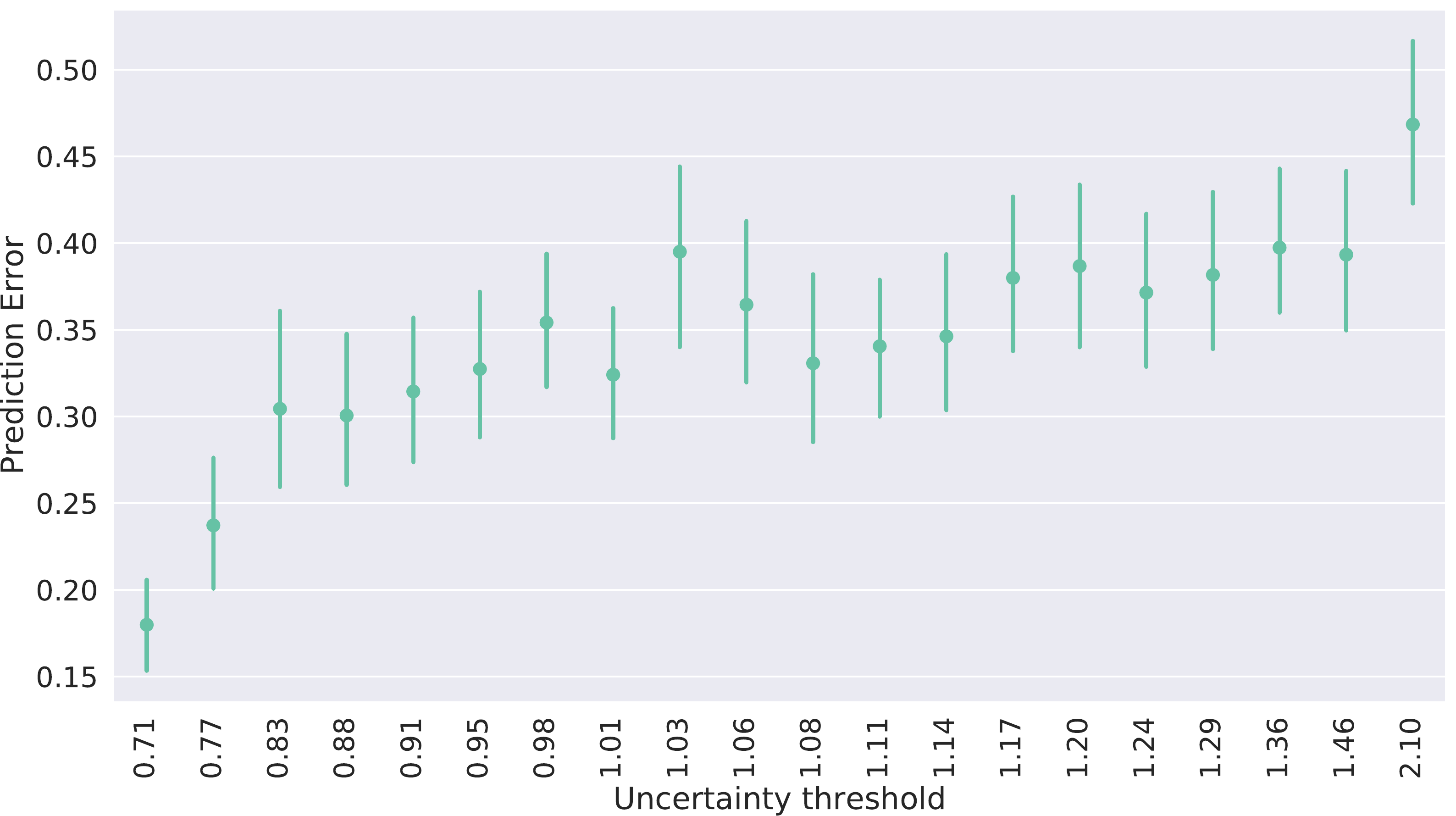}
    \caption{Mean error per uncertainty bucket for human microsomal clearance neural network model with Graph Convolution features}
    \label{fig:microsomal_nn_gc_ci}
\end{figure}
 These curves show that the featurizer and model type have a strong effect on the relationship between UQ and error.

 For human plasma protein binding HSA, which is the largest dataset with over 123,000 compounds, all calibration curves display the desired behavior: error increases as uncertainty increases and the 95 percent confidence intervals are small.
\begin{figure}[htp]
    \centering
    \includegraphics[width=\columnwidth]{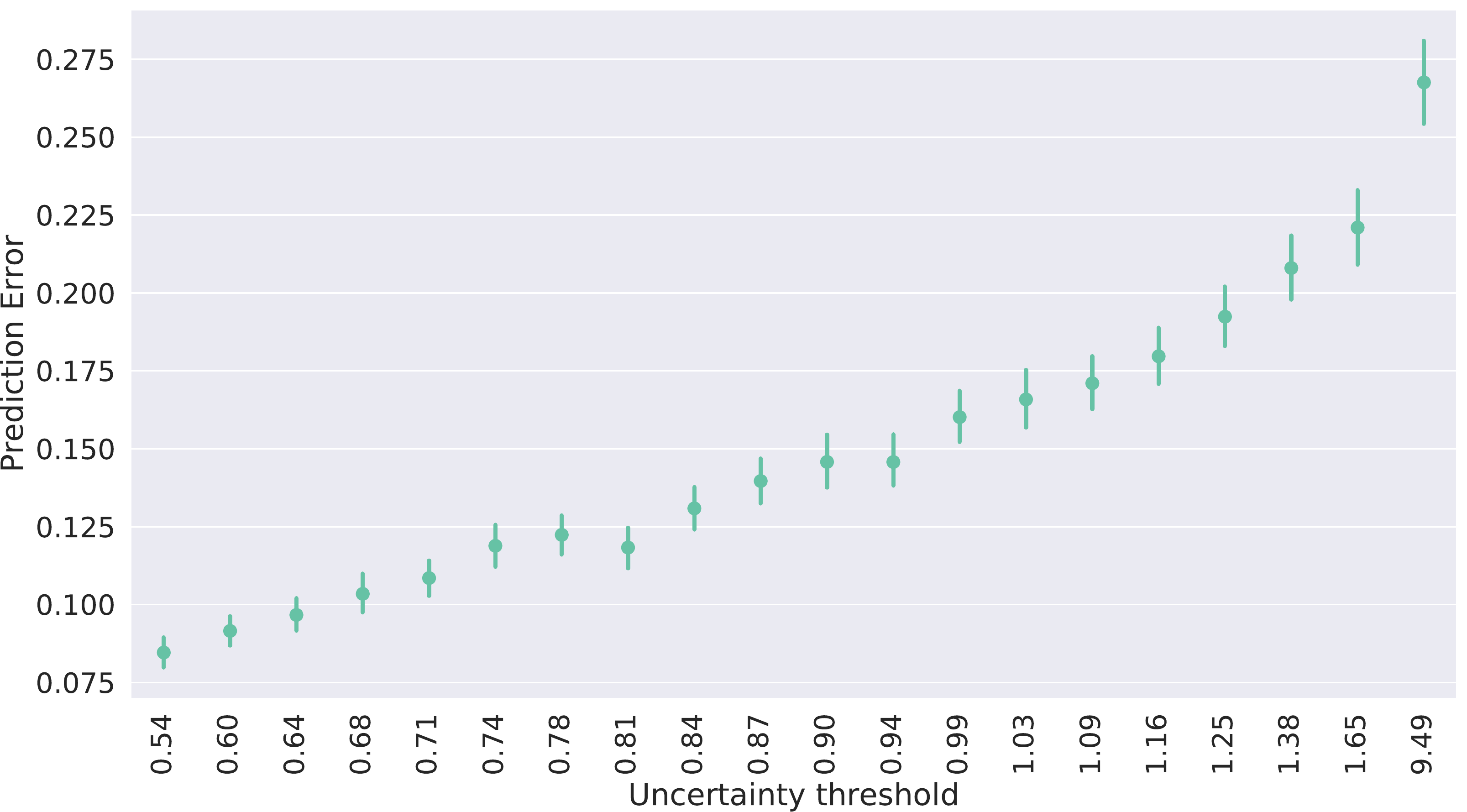}
    \caption{Mean error per uncertainty bucket for human plasma protein binding HSA neural network model with MOE features}
    \label{fig:ppb_nn_moe_ci}
\end{figure}

\begin{figure}[htp]
    \centering
    \includegraphics[width=\columnwidth]{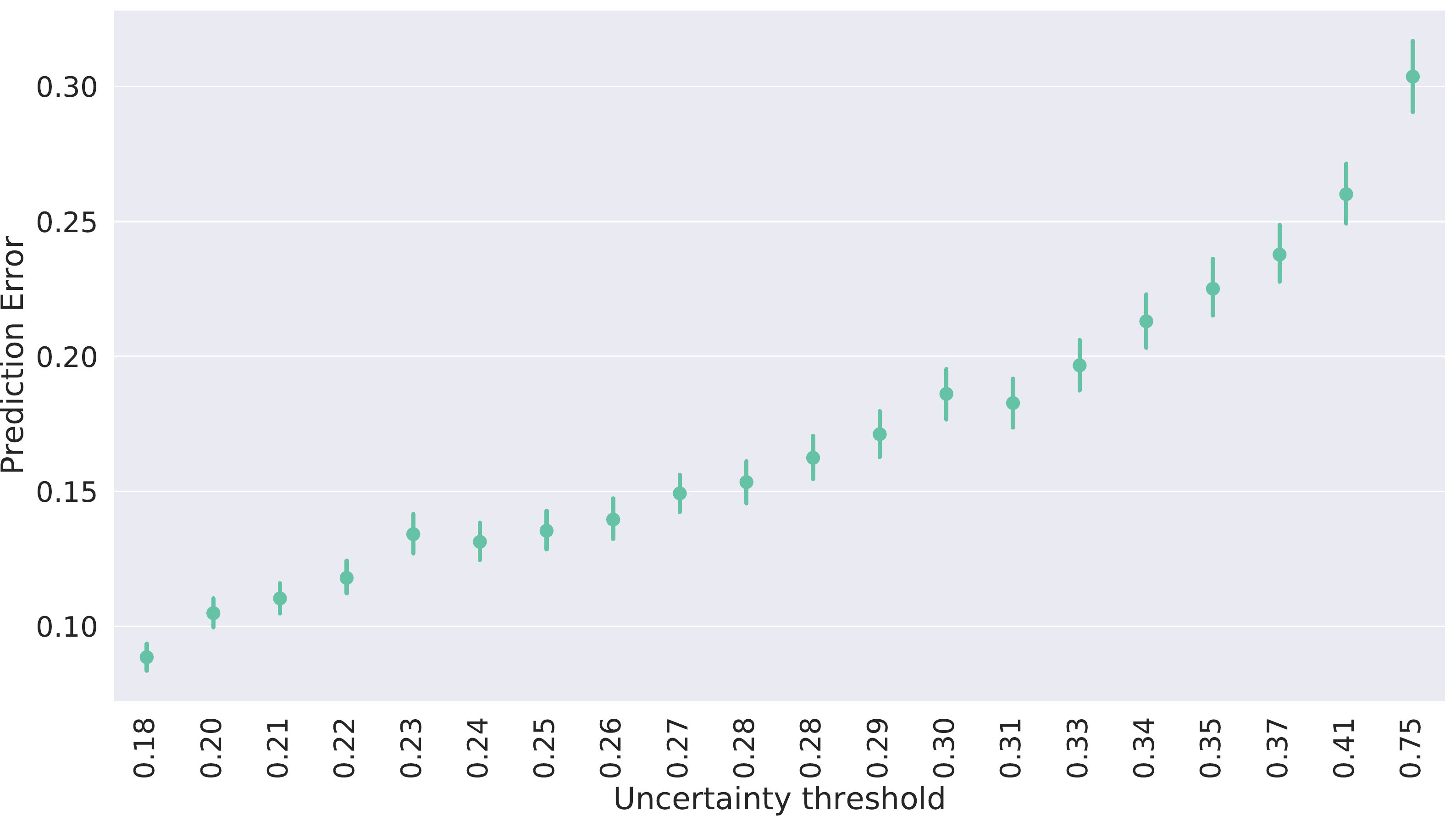}
    \caption{Mean error per uncertainty bucket for human plasma protein binding HSA random forest model with MOE features}
    \label{fig:ppb_rf_moe_ci}
\end{figure}

\begin{figure}[htp]
    \centering
    \includegraphics[width=\columnwidth]{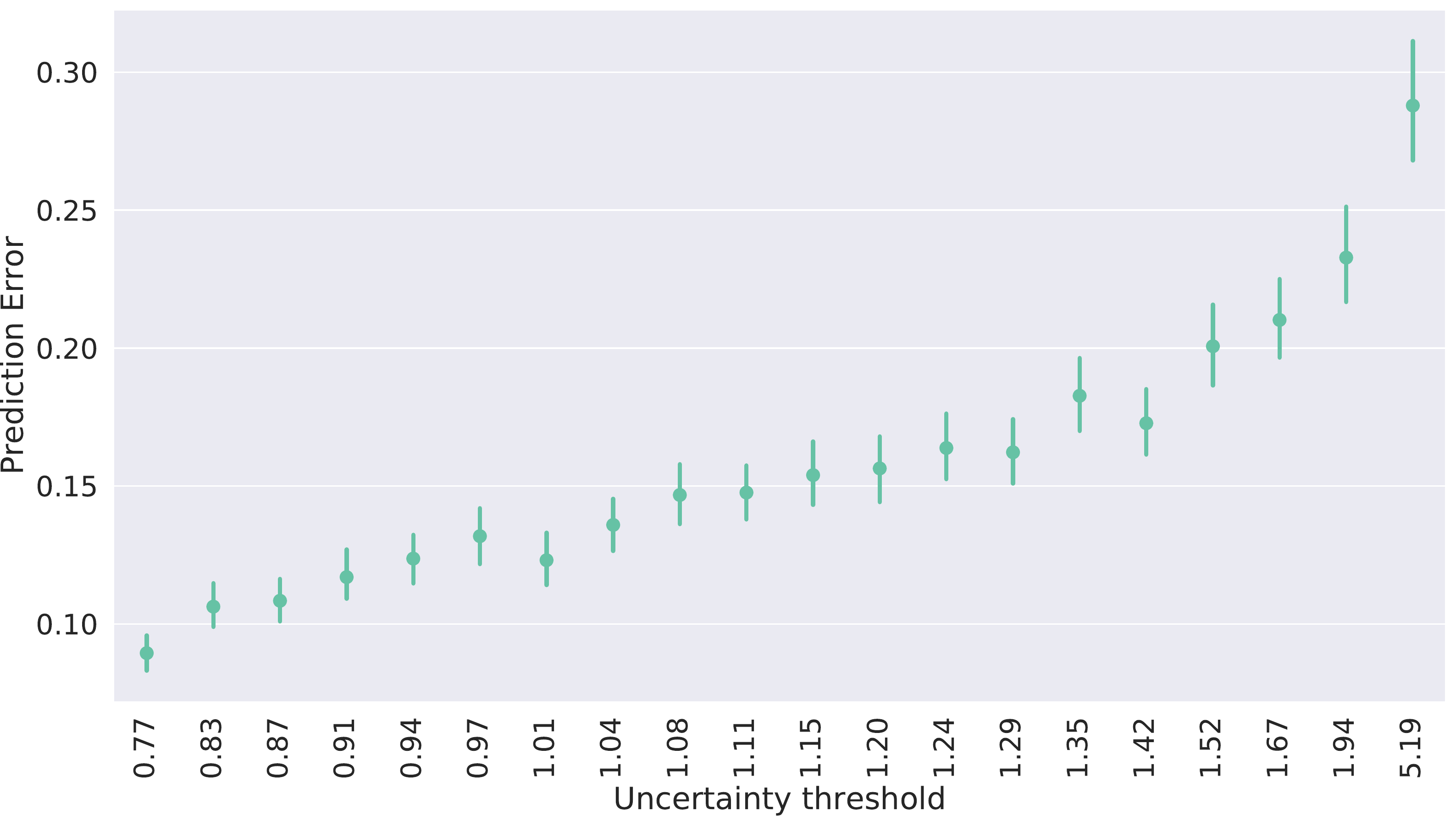}
    \caption{Mean error per uncertainty bucket for human plasma protein binding HSA neural network model with Graph Convolution features}
    \label{fig:ppb_nn_gc_ci}
\end{figure}

\subsubsection{Examining the relationship between UQ and predicted value}
Since the UQ values quantify the variation in predictions, the relationship between UQ and the predicted values were checked for evidence of a correlation by examining plotted UQ versus predicted values.

Rat plasma clearance (\textit{in vivo}) shows a somewhat negative relationship, where the variation in predictions decreases as the magnitude of the predicted value increases. We found a similar though much less pronounced trend when examining error versus predicted value, so it looks like overall the model is predicting better for compounds with higher clearance values.
\begin{figure}[htp]
    \centering
    \includegraphics[width=\columnwidth]{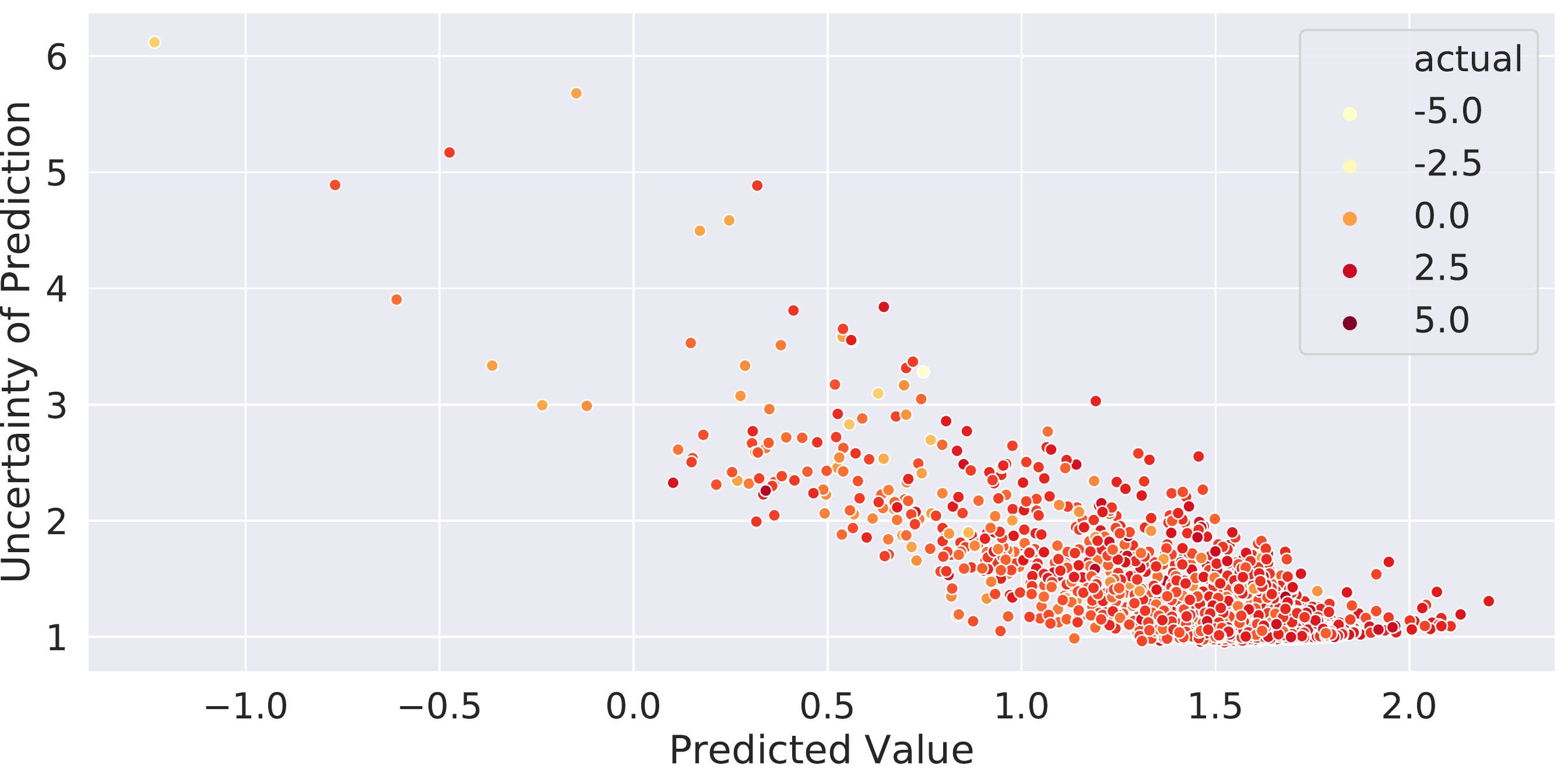}
    \caption{Uncertainty value versus Predicted for rat plasma clearance (\textit{in vivo}) neural network model with MOE features}
    \label{fig:invivo_nn_moe_std_pred}
\end{figure}

\begin{figure}[htp]
    \centering
    \includegraphics[width=\columnwidth]{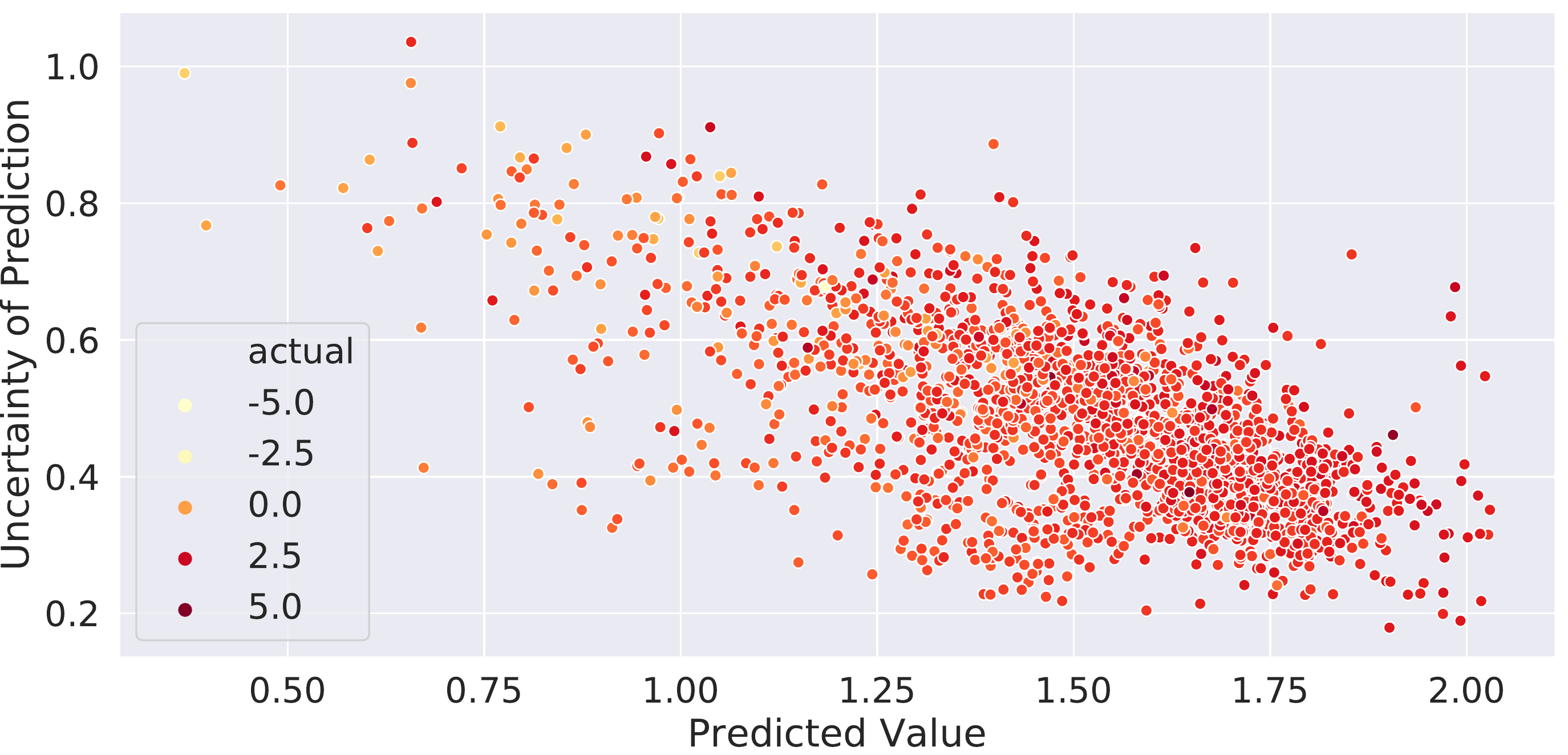}
    \caption{Uncertainty value versus Predicted for rat plasma clearance (\textit{in vivo}) random forest model with MOE features}
    \label{fig:invivo_rf_moe_std_pred}
\end{figure}

\begin{figure}[htp]
    \centering
    \includegraphics[width=\columnwidth]{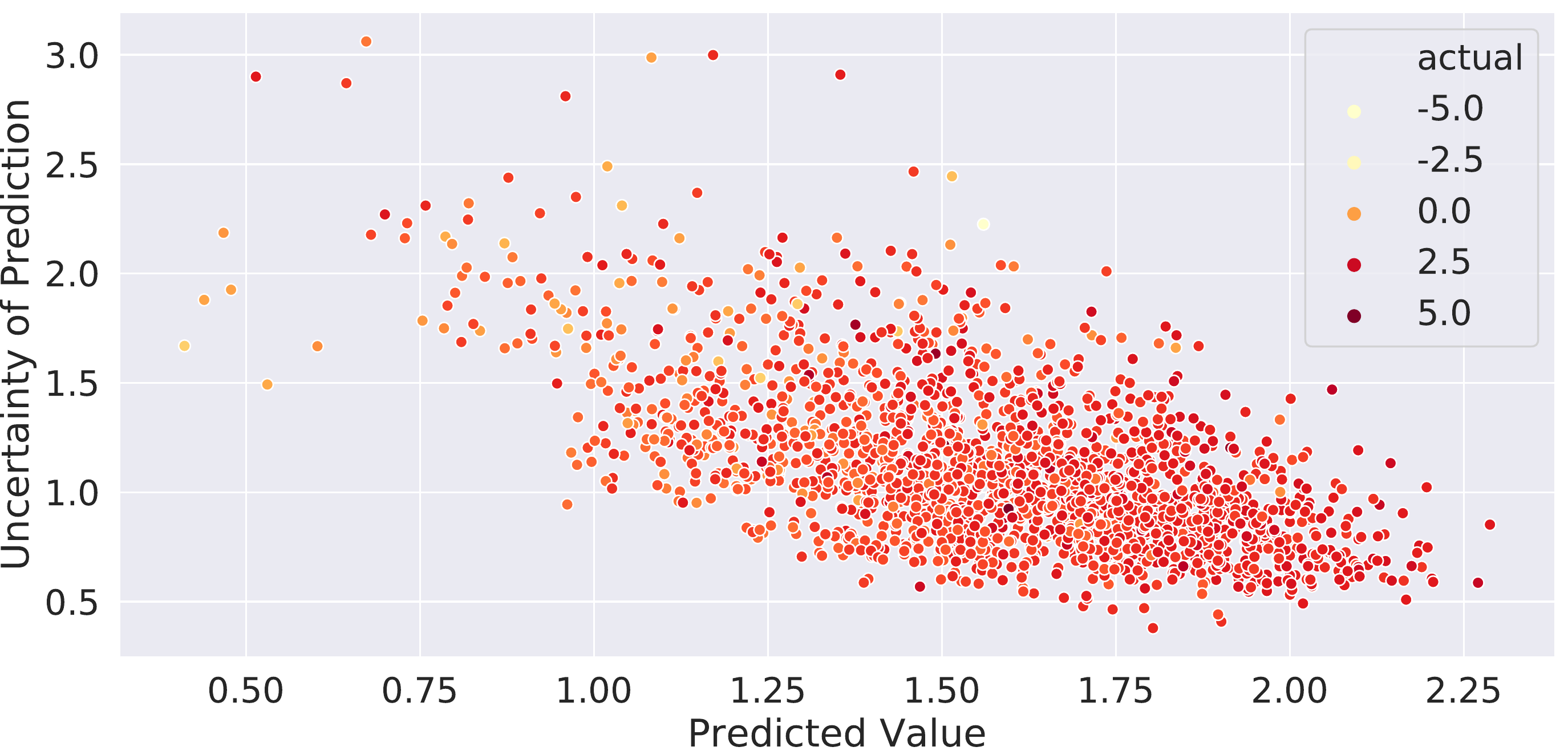}
    \caption{Uncertainty value versus Predicted for rat plasma clearance (\textit{in vivo}) neural network model with Graph Convolution features}
    \label{fig:invivo_nn_gc_std_pred}
\end{figure}

\begin{figure}[htp]
    \centering
    \includegraphics[width=\columnwidth]{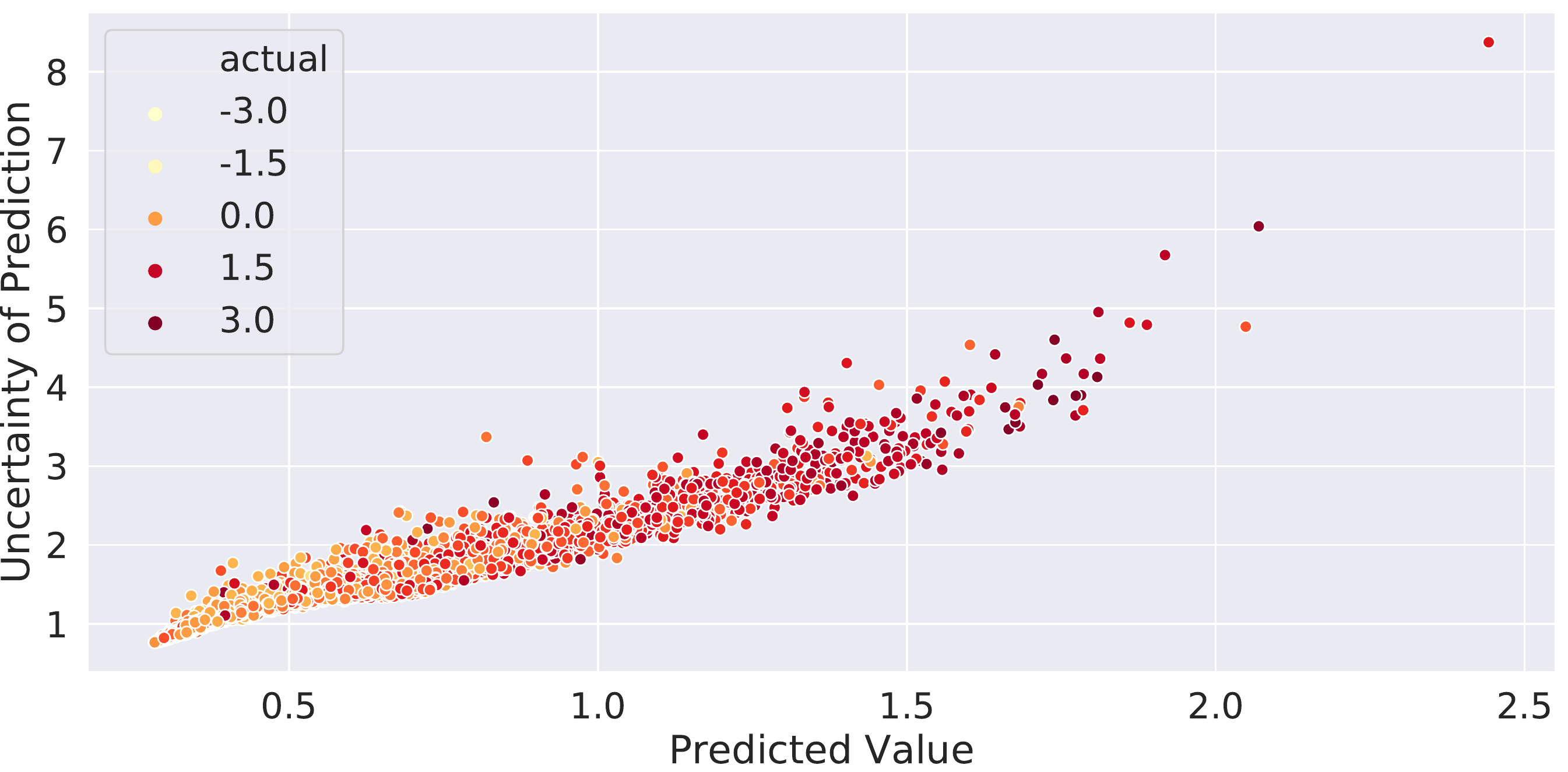}
    \caption{Uncertainty value versus Predicted for human microsomal clearance neural network model with MOE features}
    \label{fig:microsomal_nn_moe_std_pred}
\end{figure}

For human microsomal clearance, MOE feature vectors yield models where the UQ is strongly biased by the predicted value, especially for the neural network model, as seen in Figure \ref{fig:microsomal_nn_moe_std_pred}. Error versus predicted value does not show this trend, so this is likely indicating that UQ contains no real information value for this model.
This trend exists for the MOE random forest model as well, although it levels off, suggesting slightly less biased UQ values.
\begin{figure}[htp]
    \centering
    \includegraphics[width=\columnwidth]{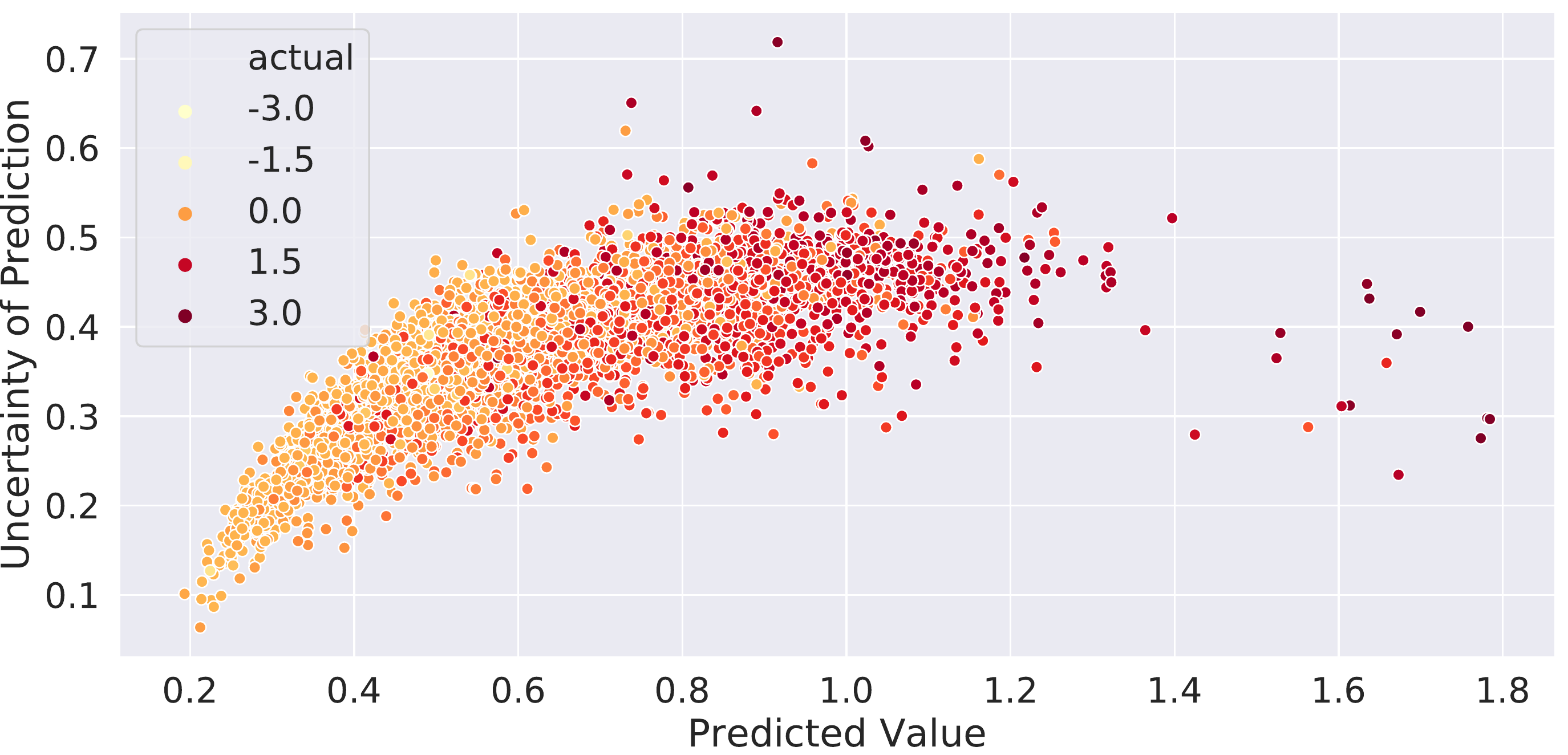}
    \caption{Uncertainty value versus Predicted for human microsomal clearance random forest model with MOE features}
    \label{fig:microsomal_rf_moe_std_pred}
\end{figure}
The graph convolution model displays a more balanced relationship between UQ and predicted value, which mirrors what we saw in the previous two sub-sections, that this model's UQ is more informative of error than the MOE models' UQ.
\begin{figure}[htp]
    \centering
    \includegraphics[width=\columnwidth]{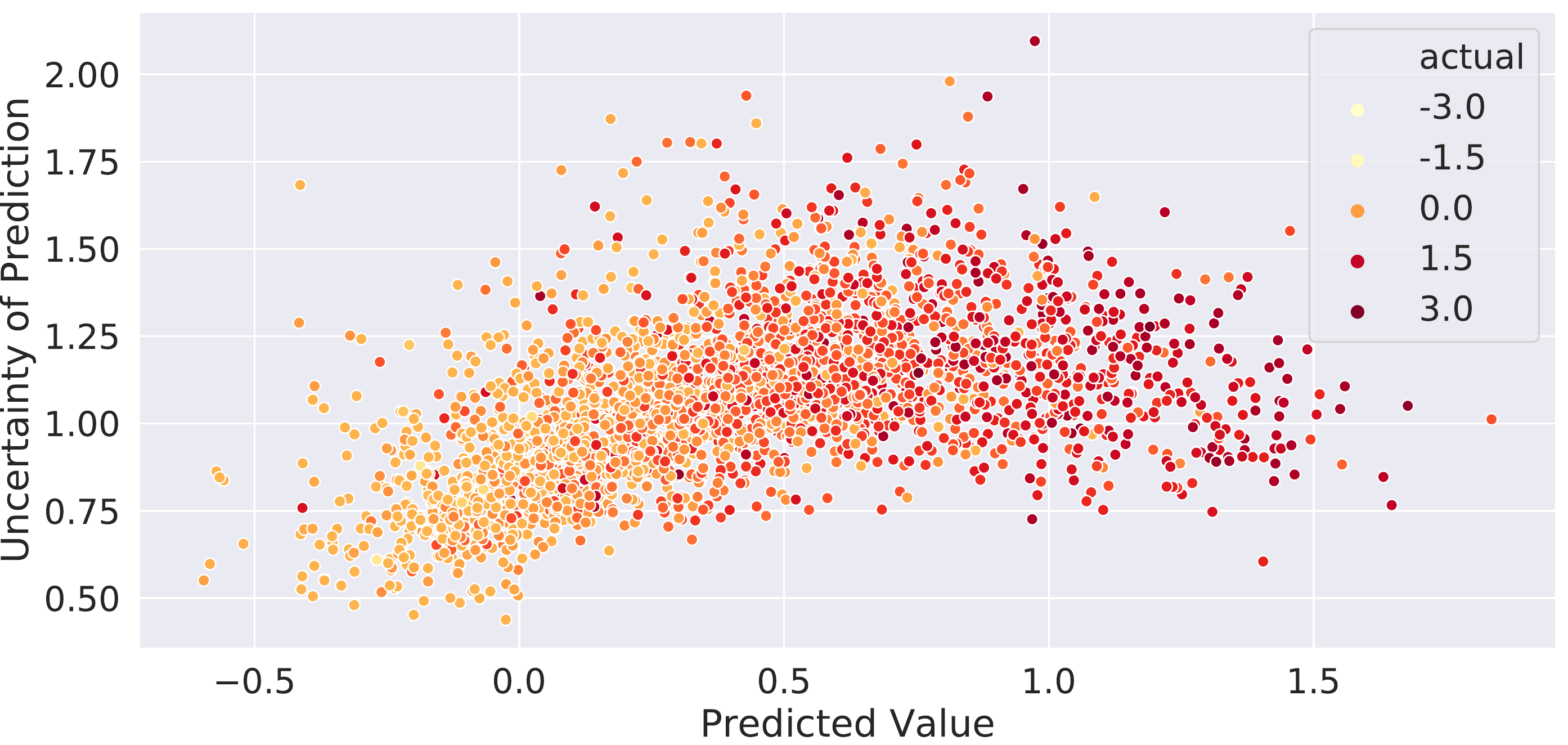}
    \caption{Uncertainty value versus Predicted for human microsomal clearance neural network model with Graph Convolution features}
    \label{fig:microsomal_nn_gc_std_pred}
\end{figure}
Human plasma protein binding HSA, which showed the best calibration curves, also shows the least correlation between UQ and predicted value. UQ has a wide range of values for all predicted values.
\begin{figure}[htp]
    \centering
    \includegraphics[width=\columnwidth]{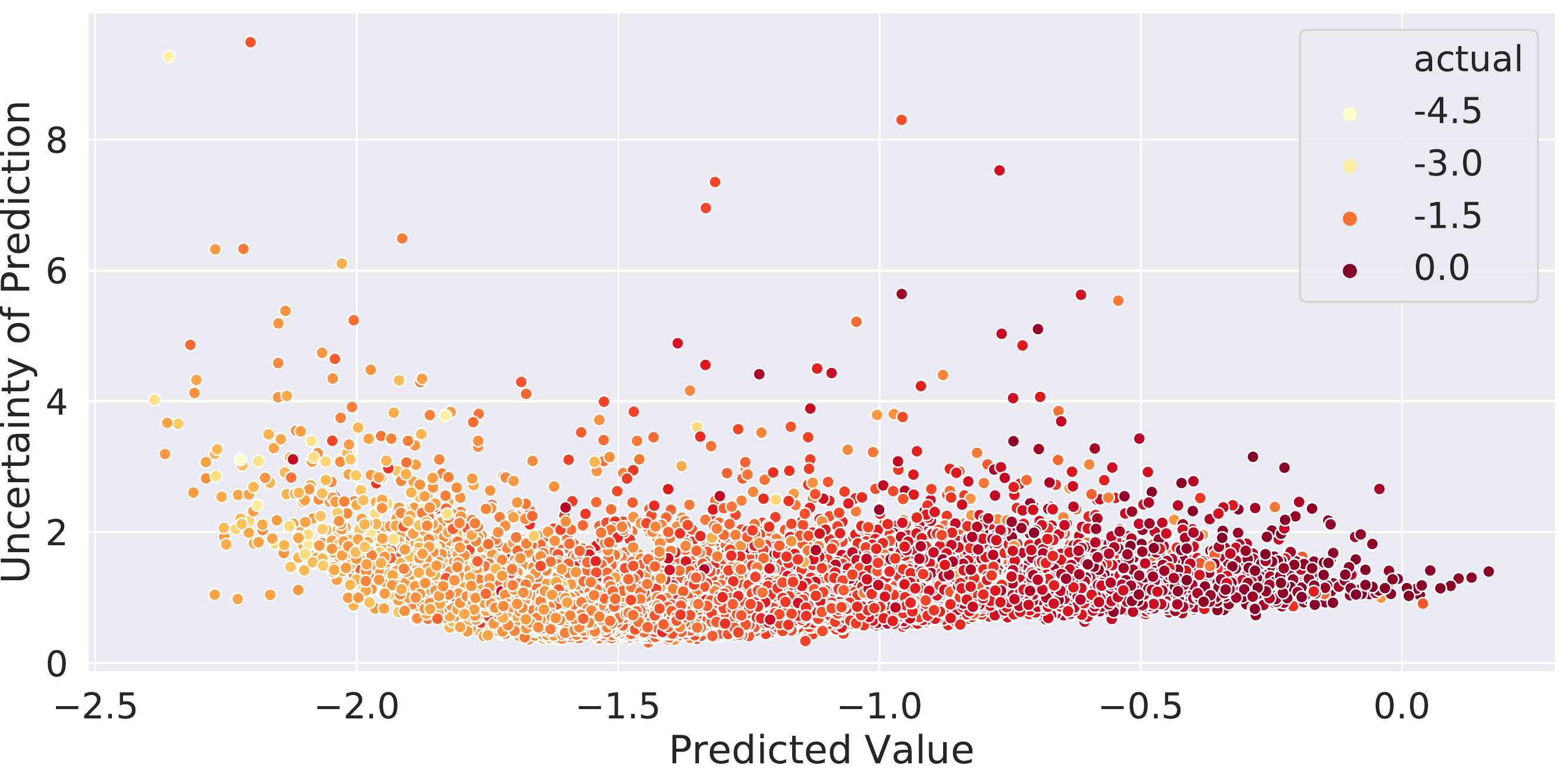}
    \caption{Uncertainty value versus Predicted for human plasma protein binding HSA neural network model with MOE features}
    \label{fig:ppb_nn_moe_std_pred}
\end{figure}

\begin{figure}[htp]
    \centering
    \includegraphics[width=\columnwidth]{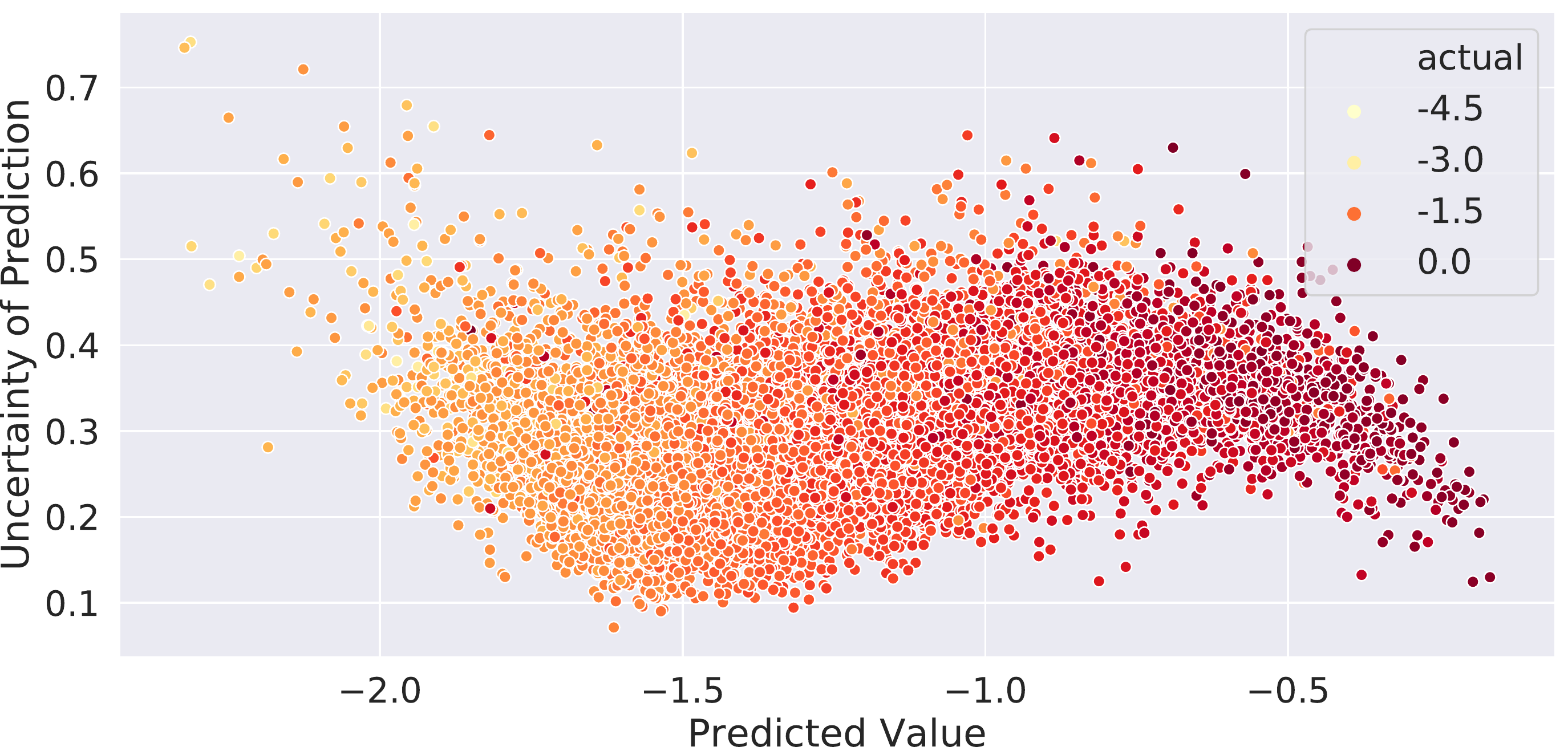}
    \caption{Uncertainty value versus Predicted for human plasma protein binding HSA random forest model with MOE features}
    \label{fig:ppb_rf_moe_std_pred}
\end{figure}

\begin{figure}[htp]
    \centering
    \includegraphics[width=\columnwidth]{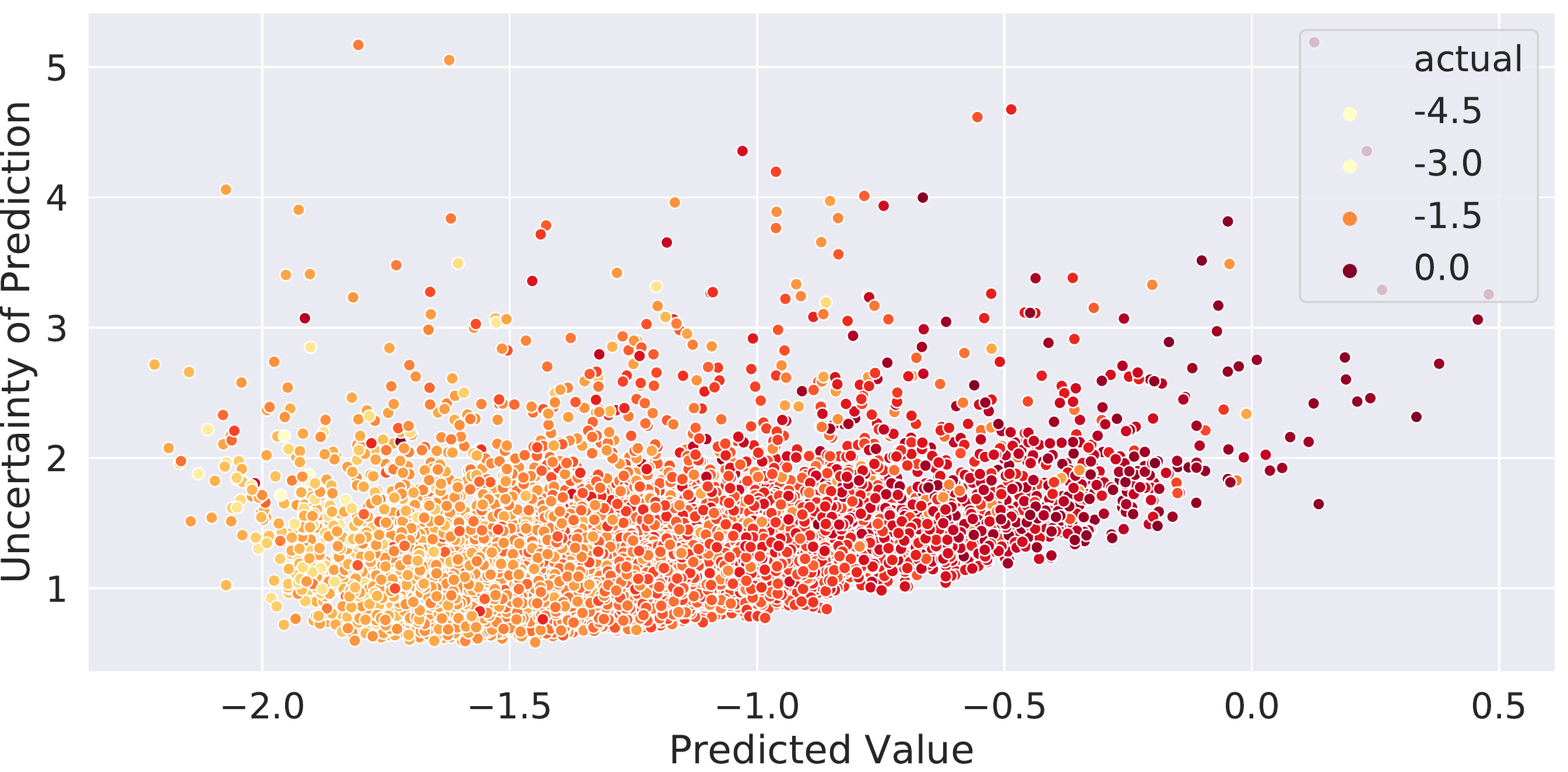}
    \caption{Uncertainty value versus Predicted for human plasma protein binding HSA neural network model with Graph Convolution features}
    \label{fig:ppb_nn_gc_std_pred}
\end{figure}
\subsubsection{Correlation between UQ and error}
\begin{figure}[htp]
    \centering
    \includegraphics[width=\columnwidth]{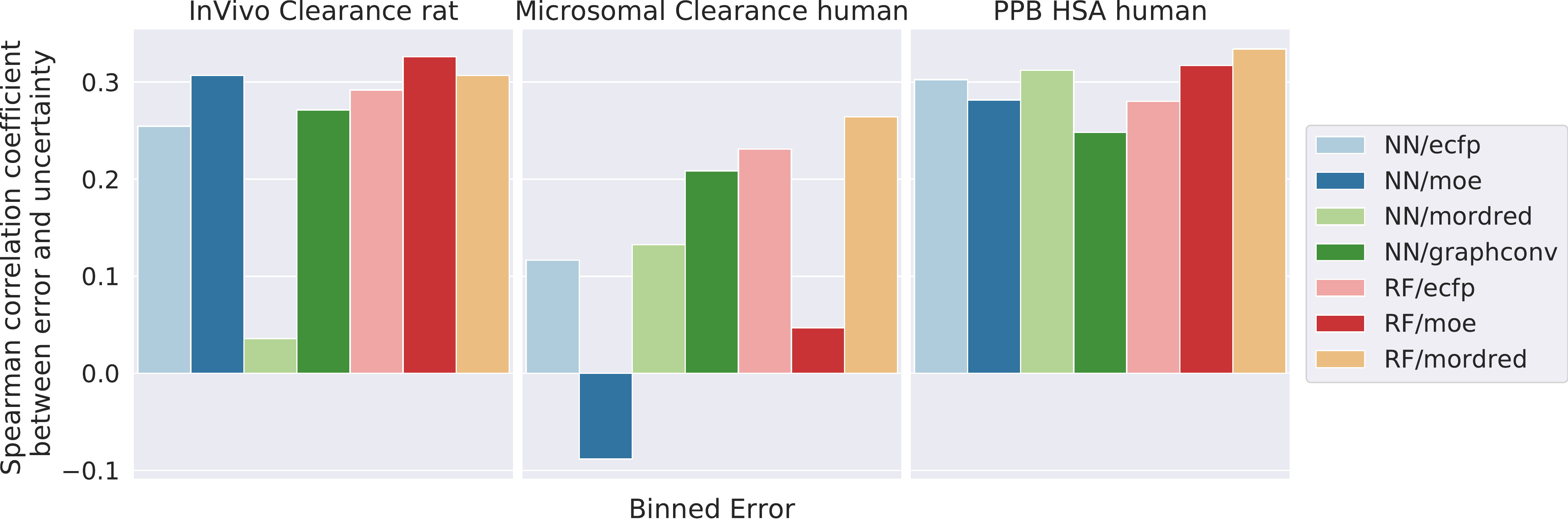}
    \caption{Spearman correlation coefficient between error and uncertainty values}
    \label{fig:error_uq_corr}
\end{figure}

While these plots provide useful methods for visualizing the behavior of uncertainty quantification, we wanted to identify a value that could summarize if we could trust a given model's UQ results. Since we want the certainty of the model to be reflected in accurate predictions, we calculated the Spearman correlation coefficient between between binned prediction error and UQ. Results are shown in Figure \ref{fig:error_uq_corr}. Correlations range from -0.088 to 0.33. While these correlations are fairly low, all p-values are $<0.05$, and all but one are $<<0.01$, so there is significance to the weak correlations identified.

\section{Discussion}

Key observations from the extensive series of model evaluations are summarized here:
\begin{itemize}
\item Neural networks generally produced more accurate models only on the larger datasets.
\item The proprietary MOE descriptors outperformed the open-source Mordred descriptors for both random forest and neural networks. Among neural network representations, graph convolutions outperformed ECFP.
\item A range of neural network architectures performed best, depending on the dataset size. Small networks appear to be  prominently featured in many datasets. 
\item Model performance generally improved as dataset size increased, suggesting the need for public dataset integration or multi-task/ transfer learning approaches.
\item Hyperparameter tuning generally improved performance, in some cases dramatically.
\item Uncertainty quantification showed a weak correlation with error, and the efficacy of using UQ to filter predictions varied considerably between datasets and model types.
\end{itemize}

The differences in prediction accuracy show that the parameters needed for \textit{in silico} drug discovery present a diverse set of data-driven modeling challenges. The extensive benchmarking suggests that there is no clear one best modeling approach for every predicted parameter.  The differences in performance show the importance of having a rigorous model building pipeline that can be readily adapted and re-applied to build parameter specific models as new data becomes available.

\section{Conclusions}

In this paper, we present the ATOM Modeling PipeLine, or \pipeline. This open-source software suite allows the user to build global and local models for a wide array of molecular properties that are needed for \textit{in silico} drug discovery. Results of extensive benchmarking on a wide variety of pharmacokinetic and safety datasets were also presented, with an exploration of the effects of different featurization and model types on model accuracy. While the datasets used for developing and testing the pipeline are not publicly available, the software used to curate data and train, evaluate, and share new models is available as open source and benefits from having been tested on a wide array of pharmaceutically-relevant parameters. Additional public datasets are included with the pipeline release to support applying reproducible training and testing protocols that enable the broader modeling community to evaluate and improve modeling approaches over time. 

\section{Disclaimer}
This document was prepared as an account of work sponsored by an agency of the United States government. Neither the United States government nor Lawrence Livermore National Security, LLC, nor any of their employees makes any warranty, expressed or implied, or assumes any legal
liability or responsibility for the accuracy, completeness, or usefulness of any information, apparatus, product, or process disclosed, or represents that its use would not infringe privately owned rights. Reference herein to any specific commercial product, process, or service by trade name, trademark, manufacturer, or otherwise does not necessarily constitute or imply its endorsement, recommendation, or favoring by the United States government or Lawrence Livermore National Security, LLC. The views and opinions of authors expressed herein do not
necessarily state or reflect those of the United States government or Lawrence Livermore National Security, LLC, and shall not be used for advertising or product endorsement purposes. 

This work was performed under the auspices of the U.S. Department of Energy by Lawrence Livermore National Laboratory under contract DE-AC52-07NA27344.
\section{Funding Sources}
This work represents a multi-institutional effort. Funding sources include: Lawrence Livermore National Laboratory internal funds; the National Nuclear Security Administration; GlaxoSmithKline, LLC; and federal funds from the National Cancer Institute, National Institutes of Health, and the Department of Health and Human Services, under Contract No. 75N91019D00024.

\section{Appendix}

\section{Benchmarking of AMPL on public datasets}
AMPL is open source and available for download at \\\texttt{http://github.com/ATOMconsortium/AMPL}. To support reproducibility of this pipeline, we provide model-building examples for three public datasets in AMPL's open source repository. These datasets include: 
\begin{itemize}
    \item Delaney et al. solubility dataset \cite{delaney}
    \item Wenzel et al. human liver microsome intrinsic clearance \cite{clearance}
    \item Drug Target Commons KCNH2 (hERG) inhibition assay \cite {herg}
\end{itemize}
Since the data from our main benchmarking experiments are proprietary, we also benchmarked AMPL on these publicly-available datasets. Results are shown below.
\begin{table*}[h!]
  \begin{adjustbox}{max width=\textwidth}
\begin{tabular}{|l|l|l|l|l|}
\hline
\textbf{Dataset} & \textbf{Model and featurizer type} & \textbf{Train set $R^2$}  & \textbf{Validation set $R^2$} & \textbf{Test set $R^2$}\\ \hline
Delaney solubility & Neural network + ECFP & 0.66 &	0.21 & 0.29 \\ \hline
Delaney solubility  & Neural network + GraphConv & 0.76 & 0.55 & 0.54 \\ \hline
Delaney solubility & Neural network + Mordred & 0.79 & 0.67 & 0.74 \\ \hline
Delaney solubility  & Random forest + ECFP & 0.91 & 0.27 & 0.37 \\ \hline
Delaney solubility  & Random forest + Mordred & 0.99 & 0.72 & 0.73 \\ \hline
Wenzel microsomal clearance & Neural network + ECFP & 0.19 &	0.07 &	0.054 \\ \hline
Wenzel microsomal clearance & Neural network + GraphConv & 0.11 & 0.064 & 0.067 \\ \hline
Wenzel microsomal clearance & Neural network + Mordred & 0.40 & 0.21 & 0.13 \\ \hline
Wenzel microsomal clearance & Random forest + ECFP & 0.90 & 0.17 & 0.21 \\ \hline
Wenzel microsomal clearance & Random forest + Mordred & 0.92 & 0.15 & 0.12 \\ \hline
KCNH2 (hERG) inhibition & Neural network + ECFP & 0.30 &	0.22 & 0.15 \\ \hline
KCNH2 (hERG) inhibition & Neural network + GraphConv & 0.28 & 0.19 & 0.18 \\ \hline
KCNH2 (hERG) inhibition & Neural network + Mordred & 0.24 & 0.20 & 0.19 \\ \hline
KCNH2 (hERG) inhibition & Random forest + ECFP & 0.90 & 0.36 & 0.38 \\ \hline
KCNH2 (hERG) inhibition & Random forest + Mordred & 0.94 & 0.39 & 0.36 \\ \hline
\end{tabular}
\end{adjustbox}

\caption{$R^2$ scores for public dataset models}
\label{tab:public}
\end{table*}

\bibliography{ampl}

\end{document}